\documentclass[%
 reprint,
 amsmath,amssymb,
 aps,
]{revtex4-2}
\usepackage{physics}
\usepackage{graphicx}% Include figure files
\usepackage{dcolumn}% Align table columns on decimal point
\usepackage{comment}
\usepackage{bm}% bold math
\usepackage{cases}
\usepackage[dvipsnames]{xcolor}
\usepackage[colorlinks=true, allcolors=blue]{hyperref}
\usepackage[T1]{fontenc} % higher quality font encoding

\newcommand{\BS}[1]{\boldsymbol{#1}}
\newcommand{\T}[1]{\text{#1}}

\begin{document}

\preprint{APS/123-QED}

\title{High-fidelity EDSR in a Si/SiGe Wiggle Well}% 

\author{Hudaiba Soomro}
\author{Minyoung Kim}
\author{Avani Vivrekar}
\author{M. A. Eriksson}
\author{Benjamin D. Woods}
\author{Mark Friesen}
\affiliation{Department of Physics, University of Wisconsin-Madison, Madison, Wisconsin 53706, USA}

\date{\today}

\begin{abstract}
Si/SiGe quantum wells that incorporate Ge concentration oscillations, known as long-period Wiggle Wells, have been shown to enhance the Dresselhaus spin-orbit coupling of conduction-band electrons.
Such intrinsic spin-orbit coupling is desirable when performing spin-qubit gate operations based on electric dipole spin resonance (EDSR) because it eliminates the need for external micromagnets. 
However, random-alloy disorder plays a key role in this materials system by spatially randomizing the valley splitting and the valley phase $\phi_{s,s}$, and it has not been fully accounted for in recent EDSR analyses.
Here, we show that alloy disorder affects EDSR in two main ways.
First, the Rabi frequency $\Omega$ acquires a dependence on the valley phase, given by $\cos\phi_{s,s}$, which causes spatial randomization of $\Omega$.
Despite this variability, we show that fast EDSR can be achieved at most locations across a given sample.
Second, a new Rabi driving mechanism emerges, enabled by valley dipoles and the hybridization of ground and excited valley states, which arise from alloy disorder and EDSR driving, respectively.
This mechanism is dominant in regions of low valley splitting.
Alloy disorder can therefore strengthen EDSR, but it can also cause gradients in $\Omega$ that lead to dephasing in the rotating frame.
We explore this problem by first locating ``sweet spots,'' where EDSR is relatively insensitive to electric-field fluctuations.
We then show that high-fidelity Rabi oscillations can be achieved in the presence of realistic charge noise. 
These results suggest that Wiggle Wells are a promising platform for high-quality, micromagnet-free gate operations.
\end{abstract}

\maketitle
\section{Introduction}
\label{sec1}
Si-based spin qubits have emerged as a promising platform for quantum computing due to their long coherence times, high-fidelity operations, scalability, and compatibility with existing fabrication techniques \cite{Loss1998,Burkard2023}.
Single- and two-qubit gate operations with fidelities above 99$\%$ have been achieved \cite{Xue2022, Madzik2022, Milss2022,Noiri2022}, with some fidelities above 99.99\% \cite{Wu2025preprint}.
Electric-dipole spin resonance (EDSR) is a common technique for implementing single-qubit gates, especially in single-spin qubits, which requires some form of spin-orbit coupling (SOC) \cite{RashbaEfros}.
For hole-spin qubits \cite{Terrazos2021}, this is normally intrinsic SOC \cite{Golovach2006, Rashba2008}.
However, this approach is seldom used for Si conduction-band electrons, because the SOC is weak \cite{Nestoklon2008}. 
In this case, it is more common to employ a synthetic SOC based on micromagnets \cite{Tokura2006}.
While highly successful, on-chip micromagnets present challenges for scalability and expose the qubits to charge-noise-induced dephasing \cite{Burkard2023}.
Spin-valley coupling also enables a version of EDSR that relies upon broken translational symmetry at the silicon interface to generate a valley-orbit coupling (e.g., at an atomic step) \cite{Corna2018,Huang2021,Cai2023}.  
However, this scheme is not well-explored, and involves a higher-order process combining valley-orbit coupling and SOC.

\begin{figure*}[t]
\centering
\includegraphics[width=0.96\textwidth]{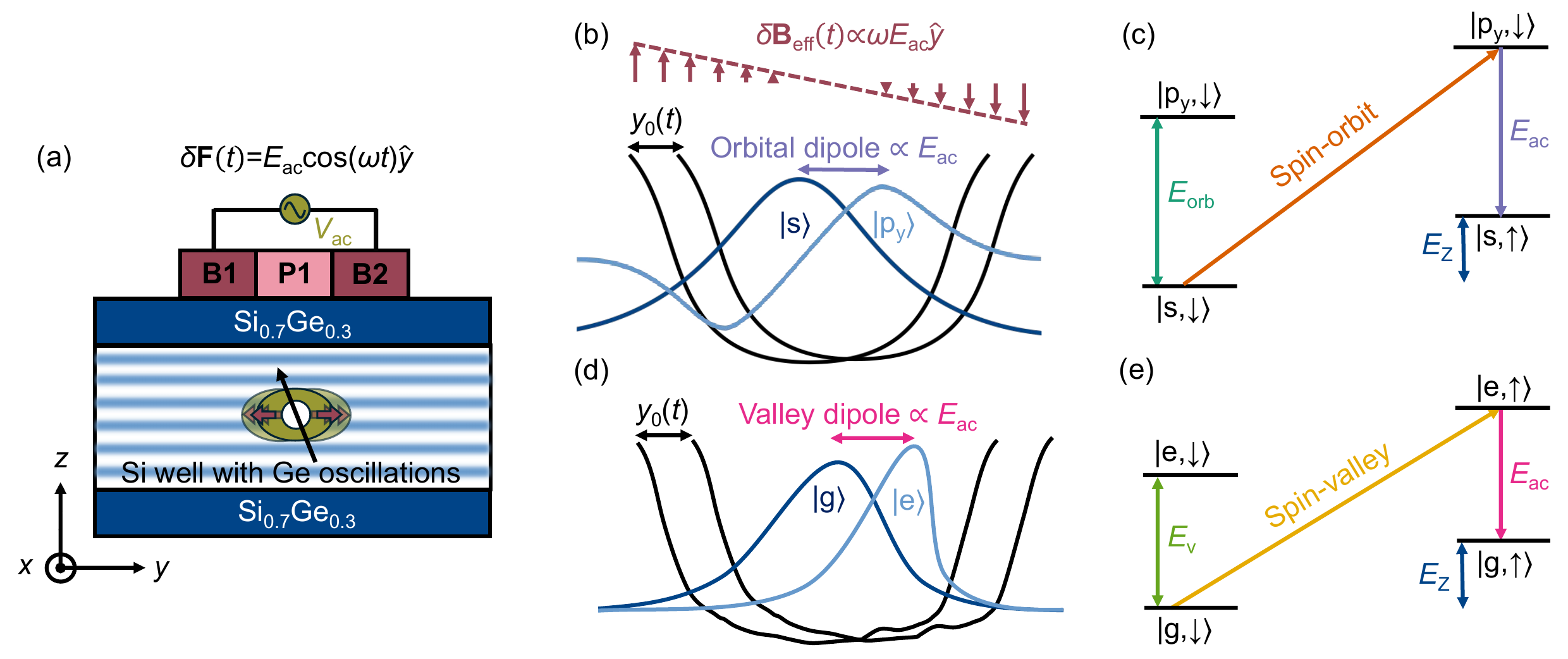}
\caption{Schematic illustration of EDSR operation in a long-period Si Wiggle Well.
(a) A quantum well is formed between SiGe barriers with Ge concentration oscillations.
A single-electron qubit is shown inside the well, under plunger gate P1, and confined between the barrier gates B1 and B2.
(Dimensions not shown to scale.)
The electron feels an in-plane ac electric field $\delta \mathbf{F}(t)$, caused by the ac voltage $V_\text{ac}$ applied to the top gates.
We also assume a vertical electric field $\mathbf{F}=F_{z}\hat{z}$ and a static magnetic field $\mathbf{B}=B\hat{x}$. 
Wiggle Well concentration oscillations of wavelength $\lambda_\text{Ge}=1.57$~nm cause a strong enhancement of the spin-orbit coupling, enabling fast EDSR.
(b) For conventional EDSR, $s$ and $p_y$ quantum-dot orbital states hybridize in response to an ac electric field, allowing the charge to oscillate along $\hat{y}$.
This response to the ac field is known as an ``orbital dipole.''
In a WW, the oscillating charge is converted to an effective oscillating magnetic field by SOC (or by spin-valley coupling).
(c) The conventional, second-order EDSR driving mechanism: spin-orbit coupling hybridizes spin and orbital states, while the ac drive hybridizes the $s$ and $p_y$ orbitals. 
(d) A new EDSR driving mechanism is enabled by alloy disorder.
In this case, the ground ($g$) and excited ($e$) valley states have a different center of charge, due to valley-orbit coupling.
The ac driving field now hybridizes these valley states, resulting in a ``valley dipole.''
(e) The valley-dipole EDSR driving mechanism: effective spin-valley coupling hybridizes the spin and valley states, while the ac drive hybridizes the $g$ and $e$ valley states.
The driving mechanisms in (c) and (e) are both present in Si WWs, yielding separate contributions to the Rabi frequency, $\Omega_\text{orb}$ and $\Omega_v$, respectively.}
\label{Fig1}
\end{figure*}

SiGe heterostructures with Ge concentration oscillations inside the quantum well, illustrated in Fig.~\ref{Fig1}(a), were originally proposed to enhance the valley-state energy splitting (or ``valley splitting,'' $E_v$) between the two low-energy valley states in quantum dots \cite{McJunkin2022, Feng2022}.
More recently, these so-called Wiggle Wells (WWs) were also predicted to enhance the SOC by 1-2 orders of magnitude compared to conventional Si quantum wells \cite{Woods2023}.
The oscillation wavelength needed for SOC enhancement, $\lambda_{\text{Ge}} \approx 1.6~\text{nm}$, has already been demonstrated experimentally \cite{McJunkin2022,Gradwohl2025}, and could therefore enable fast EDSR without micromagnets.
However, Ge forms a disordered random alloy in Si, which has not been fully accounted for in the current SOC theory, although it is known to play a prominent role in the physics of Si/SiGe quantum dots \cite{Wuetz2021,Losert2023,Lima2023b,Lima2023a,Pena2024,Marcks2025}.
Therefore, the effect of alloy disorder on EDSR implementations still remains an important open question.

In this work, we show that random-alloy disorder has two significant effects on EDSR in a WW. 
First, it causes spatial variations of the valley phase $\phi_{s,s}$, defined below, which in turn cause spatial variations of the conventional EDSR Rabi frequency, $\Omega_\text{orb}$.
Second, alloy disorder brings a new contribution to the Rabi frequency, $\Omega_v$, arising from SOC and the presence of ``valley dipoles,'' which describe the hybridization of the ground and excited valley states in response to an EDSR driving field.
It is important to note that the center positions of these valley states are shifted due to valley-orbit coupling \cite{Gamble2013}, as illustrated in Fig.~\ref{Fig1}(d); the valley-dipole ESDR driving mechanism therefore cannot occur in the absence of alloy disorder.
Importantly, we show here that $\Omega_v \propto E_v^{-2}$, which leads to spatially varying, strong enhancements of $\Omega_v$ in regions of low valley splitting.

We also explore a second topic related to spatially fluctuating device parameters. 
Electric field fluctuations (or ``charge noise'') can cause unintended shifts of the dot position, with potential consequences for Rabi-mediated $X(\theta)$ gate operations.
First, small spatial fluctuations of the $g$-factor can occur \cite{WoodsGFactor,Woods2025b}, yielding fluctuations of the spin resonance frequency; this effect is relatively weak and we ignore it here.
We focus instead on a much stronger effect arising from spatial variations of the total Rabi frequency $\Omega=\Omega_\text{orb}+\Omega_v$, which cause random under- or overrotations of $X(\theta)$, and ultimately limit the Rabi dephasing time $T_{2,\text{Rabi}}$.
Below, we characterize this decoherence mechanism by computing spatial maps of $T_{2,\text{Rabi}}$, and the corresponding quality factor $Q_\text{Rabi} = \Omega T_{2,\text{Rabi}}$, for typical realizations of random-alloy disorder.
Importantly, we obtain high $Q_\text{Rabi}$ values for most quantum dot locations across the map.
Small $Q_\text{Rabi}$ values are typically observed only in regions of low valley splitting, where $\Omega_v$ is large, but varies rapidly in space.
Additionally, we observe ``sweet spots'' with especially high $Q_\text{Rabi}$ values, where $T_{2,\text{Rabi}}$ is first-order insensitive to electric-field fluctuations in all three spatial directions.
The main conclusions of our work therefore suggest that (i) WWs provide a desirable platform for single-spin qubits, allowing high-fidelity, micromagnet-free EDSR gate operations; however, (ii) the dots should be tuned away from regions of low valley splitting.

The paper is organized as follows. 
In Sec.~\ref{sec: model}, we describe our theoretical model for EDSR in the WW architecture. 
In Sec.~\ref{sec: EDSR theory}, we derive the orbital- and valley-dipole contributions to the EDSR Rabi frequency. 
We also compute the statistical distribution functions for valley-coupling matrix elements that contribute to the orbital and valley EDSR mechanisms, with details given in the Appendices. 
In Sec.~\ref{SpatialFlucsXY}, we present results for the spatially varying Rabi frequencies caused by alloy disorder. 
We also present results for dephasing times, dephasing rates, and quality factors, arising from in-plane charge noise, and we identify sweet spots, where charge-noise effects are minimized. 
In Sec.~\ref{LongitudinalNoise}, we account for the out-of-plane component of the electric-field noise and we demonstrate the robustness of the sweet spots to this noise contribution. 
Finally, in Sec.~\ref{conclusion}, we conclude.

\section{Model}
\label{sec: model}
The bulk band structure of Si is six-fold degenerate, with conduction-band valleys located near the $X$, $Y$, and $Z$ points of the Brillouin zone \cite{Zwanenburg2013}. 
Here, we consider the Si/SiGe WW heterostructure shown schematically in Fig.~\ref{Fig1}(a).
The strain-relaxed SiGe barrier layers induce tensile strain in the quantum well that partially lifts the valley degeneracy, leaving only two low-energy, degenerate $Z$ valleys at the bottom of the conduction band \cite{Schaffler1997}.
We label these valley basis states $\ket{+z}$ and $\ket{-z}$.

We consider the Hamiltonian $H = H_0 + H_v$, expressed in terms of its intravalley ($H_0$) and intervalley ($H_v$) parts.
The intravalley Hamiltonian takes the form
\begin{equation}
    H_{0}=\left[\frac{\hbar^2}{2m_{t}}(\hat{k}_{x}^2+\hat{k}_{y}^2)+\frac{\hbar^2}{2m_{l}}\hat{k}_{z}^2+V(\mathbf{r},t)\right] +
    \frac{E_Z}{2} \sigma_x ,
    \label{Eq1b}
\end{equation}
where $\hat{k}_{j}=-i\partial_{j}$, and $m_{l}=0.91m_0$ and $m_{t}=0.19m_0$ are the longitudinal and transverse effective masses, respectively.
The Zeeman energy is given by $E_Z = g \mu_B B$, where $g$ is the Land\'{e} $g$-factor and ${\sigma}_{j}$ are Pauli spin operators with $j\in\{x,y,z\}$.
Here, we take the magnetic field $B$ to be oriented along the $[100]$ crystallographic ($\hat x$) axis. 
The quantum dot confinement potential is given by
\begin{multline}
V(\mathbf{r},t)=\frac{m_{t}\omega_{0}^{2}}{2}
[ (x-x_d)^2+(y-y_d)^2 ]+eE_{\text{ac}}y\cos(\omega t) \\
+V_\text{conf}(z)
+V_\text{dis}(\mathbf{r})+eF_zz,
\label{Pot}
\end{multline}
where $\omega_{0}$ characterizes a two-dimensional (2D) harmonic oscillator confinement potential, induced by the plunger and barrier gates [P1, B1, and B2 in Fig.~\ref{Fig1}(a)], and $E_{\text{ac}}$ describes the strength of the in-plane ac electric field that drives EDSR at the oscillation frequency $\omega$.
$V_\text{conf}(z)$ describes the quantum-well confinement potential, as defined in Appendix~\ref{sec: numerical details}, $V_\text{dis}(\mathbf{r})$ accounts for the random Ge fluctuations (with $\langle V_\text{dis} \rangle = 0$), as described in Appendix~\ref{appendix_disorder}, and the vertical electric field $F_z$ is provided by the top gates.
The coordinates $(x_d,y_d)$ describe the center position of the dot, which will be rastered across the sample, as described below, to obtain disorder maps.
In Eq.~(\ref{Eq1b}), for simplicity, we have neglected the magnetic vector potential, whose presence imparts a correction to the effective $g$-factor \cite{WoodsGFactor}; since this correction is weak compared to the overall effect of alloy disorder, we simply ignore it here, setting $g=2$.

The intervalley Hamiltonian is given by
\begin{equation}
     H_v=\left[V(\mathbf{r})e^{-i2k_{0}z}+\beta(\hat{k}_{x}\sigma_{x}-\hat{k}_{y}\sigma_{y})e^{i2k_{1}z}\right] \tau_{-} + h.c., \label{HV}
\end{equation}
where $2k_0\hat{z}$ and $-2k_1\hat{z}$ are the intra- and inter-Brillouin-zone $k$-vectors, respectively, that couple the $\ket{+z}$ and $\ket{-z}$ valleys \cite{Woods2024,WoodsGFactor,Thayil2024}, and $\tau_{\pm} = \ket{\pm z} \bra{\mp z}$ are the valley raising/lowering operators.
The first term in Eq.~(\ref{HV}) describes the spin-preserving intervalley coupling, which gives rise to valley splitting and determines the valley eigenstates, which are formed of linear combinations of $\ket{+z}$ and $\ket{-z}$.
It is important to note the rapidly oscillating factors $e^{\pm i2k_{0}z}$ in Eq.~(\ref{HV}), which ensure that only features with wavevectors $k_z \approx  \pm 2 k_0$  in $V(\mathbf{r})$ contribute significantly to the intervalley coupling.
(This is known as the $2k_0$ theory \cite{Friesen2007_2}.)
Such feature sizes are atomic in scale, with $2\pi/2k_0 \approx 0.32~\text{nm}$. 
Since most realistic quantum wells have somewhat larger interface widths of order $w \approx 1~\text{nm}$ \cite{Wuetz2021} (see Appendix~\ref{sec: numerical details}), the main contribution to the valley coupling is therefore expected to arise from alloy disorder, which naturally contains atomic length scales \cite{Losert2023}. 
(Again, we do not consider heterostructures like short-period WWs here, which would produce a deterministic valley coupling.)
Hence, the only term in $V(\mathbf{r})$ that contributes directly to Eq.~(\ref{HV}) is $V_\text{dis}(\mathbf{r})$.
The second term in Eq.~(\ref{HV}) describes the intervalley Dresselhaus SOC \cite{Woods2023}, where $\beta = 8.2~\text{meV}\cdot\text{nm}$ \cite{WoodsGFactor}.
In this case, only features with $k_z \approx  \pm 2 k_1$ can generate SOC.
This larger wavelength, $\lambda_\text{Ge}=2\pi/2k_1 \approx 1.57~\text{nm}$, defines the long-period WW. 
In Eq.~(\ref{HV}), we note that the enhancement of the SOC arises indirectly, through oscillations of the electron wave function, as shown below.
We also note that Rashba SOC is present in this system; however, it has a much smaller magnitude than Dresselhaus SOC \cite{Woods2023}, and we ignore it here.

\section{EDSR Theory}
\label{sec: EDSR theory}
It is convenient to project the full Hamiltonian $H$ onto an effective, low-energy Hamiltonian, which can then be used to compute the EDSR Rabi frequency. 
We accomplish this, below, by first projecting the Hamiltonian onto a suitable, low-dimensional basis set, followed by a sequence of Schrieffer-Wolff transformations. 
In this scheme, alloy disorder serves mainly as a weak intervalley coupling mechanism, and is therefore treated perturbatively, within the Schrieffer-Wolff formalism.

The initial basis states in this derivation are obtained by solving the intravalley Hamiltonian $H_0$ in the absence of alloy disorder (i.e., $V_\text{dis}=0$).
These eigenstates are denoted by their quantum numbers, $\{\ket{\nu,n_x, n_y,\tau,\sigma}\}$, where $\nu$, $n_x$, and $n_y$ are spatial orbital indices, and $\tau \!\in\! \{+z,-z\}$ and $\sigma \!=\! \{\uparrow,\downarrow\}$ refer to valley and spin states, respectively. 
The separability of the potential allows us to decompose the spatial orbitals as 
\begin{equation}
    \bra{\mathbf{r}} \ket{\nu,n_x,n_y} = \varphi_{\nu}(z) \chi_{n_x,n_y}(x,y;t).
\end{equation}
Here, $\varphi_\nu$ is a quantum-well subband envelope function, satisfying the equation
\begin{equation}
    \left(-\frac{\hbar^2}{2m_{l}}\frac{d^2}{d z^2} + V_\text{conf}(z)+eF_zz \right)\varphi_\nu(z)
= \varepsilon_\nu\,\varphi_\nu(z) , \label{SubbandSchro}
\end{equation}
and $\varepsilon_{\nu}$ is the subband energy.
Similarly, $\chi_{n_x,n_y}$ is an in-plane harmonic-oscillator orbital wave function, satisfying the equation
\begin{multline}
\bigg[\frac{\hbar^2}{2m_{t}}(\hat{k}_{x}^2+\hat{k}_{y}^2) +
\frac{m_{t}\omega_{0}^{2}}{2} |{\bm\rho}-{\bm\rho}_d|^2 \\ 
+ eE_{\text{ac}}y\cos(\omega t)
\bigg] \chi_{n_x,n_y}({\bm\rho}; t) \\
= \hbar \omega_0(n_x +n_y +1)\chi_{n_x,n_y}({\bm\rho}; t) .
\label{eq:chi}
\end{multline}
In some cases, for clarity, we will explicitly include the dot location in our wave-function notation: $\chi_{n_x,n_y}({\bm\rho};{\bm\rho}_d)$, where ${\bm\rho}=(x,y)$ and ${\bm\rho}_d=(x_d,y_d)$.
We solve Eq.~(\ref{eq:chi}) at a fixed time $t$, after performing a ``moving-dot'' transformation, defined by $U_\text{osc}=e^{-i\hat{k}_{y}y_{0}(t)}$ (see \cite{Rashba2008} and Appendix~\ref{appendix_a} for details), where $y_0(t) = (eE_\text{ac}/m_t \omega_0^2) \cos(\omega t)$ describes the instantaneous center position of the oscillating harmonic confinement potential.
This transformation is performed without approximations, and has the effect of eliminating the driving term in Eq.~(\ref{eq:chi});
however, the time dependence is then transferred to the wave function through the definition $\chi_{n_x,n_y}(x,y;t) = \chi_{n_x,n_y}(x,y\! -\! y_{0}(t))$.

Only a small subset of spatial orbitals $\ket{\nu,n_x,n_y}$ are needed to capture the main EDSR effects. 
To see this, we first note that the subband energy splitting is the largest energy scale in the problem (by far). 
We may therefore limit our analysis to the lowest-energy subband, $\nu = 0$.
Consequently, the potential terms $eF_zz$ and $V_\text{conf}(z)$ no longer play a role.
This could seem surprising, since the vertical field $F_z$ sets the scale in many SOC problems; in the present case, the WW takes over the role 
of setting the spin-orbit strength \cite{Woods2023}.
Second, we note that the ac driving term in Eq.~(\ref{Pot}) couples the low-energy harmonic-oscillator $s$ mode ($n_x=n_y=0$) to the $p_y$ mode ($n_x=0$, $n_y=1$), only. 
We can therefore limit our analysis to just these two orbitals.
Henceforth, we drop $n_x$ from our notation, $\chi_{n_x,n_y} \rightarrow \chi_{n_y}$
Including the spin and valley degrees of freedom then gives a total of eight basis states, with the Hamiltonian 
\begin{equation}
    H = \begin{pmatrix}
        H_{s,s} & H_{p_y,s}^\dagger \\
        H_{p_y,s} & H_{p_y,p_y}
    \end{pmatrix}. \label{HamMatrix}
\end{equation}
Here, the intraorbital blocks are given by
\begin{align}
    H_{s,s} =&~ \frac{E_Z}{2} \sigma_x +\Delta_{s,s}(t) \tau_- + \Delta_{s,s}^*(t) \tau_+ 
    , \\
    H_{p_y,p_y} =&~ \frac{E_Z}{2} \sigma_x + \hbar \omega_0 +\Delta_{p_y,p_y}(t) \tau_- + \Delta_{p_y,p_y}^*(t) \tau_+,
\end{align}
and the interorbital blocks are given by
\begin{multline}
    H_{p_y,s} =
    ieE_\text{ac}\omega(m\omega_0^22\ell_t)^{-1} \sin(\omega t) + \Delta_{p_y,s}(t) \tau_- \\ + \Delta_{p_y,s}^*(t) \tau_+ + i\left(\beta_{0,0} \tau_- + \beta_{0,0}^* \tau_+ \right)(2\ell_t)^{-1} \sigma_y,
    \label{eq:Hpys}
\end{multline}
where the first term arises from the $-i\hbar U_\text{osc}^{\dagger}(dU_\text{osc}/dt)$ dynamical correction to the moving-dot transformation, described above, and
$\ell_t = \sqrt{\hbar/2m_t \omega_0}$ is the transverse radius of the dot, set by the harmonic confinement potential.
We define the intervalley matrix element as
\begin{multline}
    \Delta_{n_y,n_y^\prime}(t) = \\
    \int |\varphi_0(z)|^2 
    V_\text{dis}(\mathbf{r}) e^{-i2k_0z}
    \chi^*_{n_y}({\bm\rho};t) \chi_{n_y^\prime}({\bm\rho};t) \, d^3r , \label{Deltann}
\end{multline}
where we have explicitly replaced the full potential $V$ with the alloy-disorder potential $V_\text{dis}$, as explained above.
The form of the time-dependence in $\Delta_{n_y,n_y^\prime}(t)$ is derived in Appendix~\ref{appendix_a}. 
The valley coupling $\Delta_{n_y,n_y^\prime}$ plays an important role for silicon qubits.
It is a complex-valued quantity with a phase defined by $\Delta_{n_y,n_y^\prime} = |\Delta_{n_y,n_y^\prime}| \exp 
\bigl(i\phi_{n_y,n_y^\prime}\bigr)$.
The special case of $\Delta_{s,s}$ is typically referred to as \emph{the} valley coupling, and $\phi_{s,s}$ as \emph{the} valley phase; similarly, $E_v=2|\Delta_{s,s}|$ is \emph{the} valley splitting.
We note that the coupling parameters $|\Delta_{n_y,n_y^\prime}|$ and $\phi_{n_y,n_y^\prime}$ implicitly depend on the dot position $(x_d,y_d)$, as illustrated below.

The SOC matrix element in Eq.~(\ref{eq:Hpys}) is defined as
\begin{equation}
    \beta_{0,0} =  \beta \int    |\varphi_0(z)|^2 e^{i 2k_1 z} \, dz ,
    \label{beta00}
\end{equation} 
which picks out the wavevector $-2k_1$ in the subband envelope function.
We see that Eq.~(\ref{beta00}) does not include disorder effects, as consistent with our previous discussion, since the vertical confinement and $\varphi_0(z)$ are only weakly affected by disorder.
On the other hand, SOC is strongly enhanced when $-2k_1$ oscillations are generated by the WW confinement potential, since these oscillations are naturally reflected in $\varphi_0(z)$.
Similar to $\Delta_{n_y,{n_y}^\prime}$, $\beta_{0,0}$ is a complex quantity, with a ``spin-orbit phase'' $\phi_\beta$ defined by $\beta_{0,0} = |\beta_{0,0}|e^{i\phi_\beta}$. 
Throughout this work, we redefine the valley phase such that $\phi_{n_y,n_y^\prime}^\prime = \phi_{n_y,n_y^\prime} - \phi_\beta $.
To a good approximation, the SOC phase $\phi_\beta$ then appears as a \emph{global} phase shift,  independent of $(x_d,y_d)$.
To simplify our notation, we finally make the gauge choice $\phi_\beta=0$, and drop the prime notation on $\phi_{n_y,n_y^\prime}$, for brevity.

We now reduce the 8-dimensional (8D) Hamiltonian in Eq.~(\ref{HamMatrix}) to an effective 2D spin Hamiltonian by means of Schrieffer-Wolff transformations \cite{ Winkler2003}, as detailed in Appendix \ref{appendix_b}.
These approximations make use of the energy hierarchy $\hbar\omega_0 \gg E_v \gg E_Z$.
The first transformation eliminates the excited  $p_y$ orbital state in favor of the low-energy $s$ orbital. 
The second transformation eliminates the excited valley state in favor of the ground valley. 
[The result of these two transformations is given in Eq.~(\ref{EqB5}).]
We also apply a rotating wave approximation, under the resonant driving condition $\hbar\omega = E_Z$, to obtain the full Rabi frequency
\begin{equation}
    \Omega = \Omega_{\T{orb}} + \Omega_v . \label{Omega}
\end{equation}

The first term in Eq.~(\ref{Omega}) involves orbital dipoles, which occur when the ac electric field virtually hybridizes the $p_y$ orbital with the $s$ orbital, as illustrated in Fig.~\ref{Fig1}(c).
This contribution to the Rabi frequency is given by
\begin{equation}
    \Omega_{\T{orb}} = \frac{ e E_{\T{ac}} |\beta_{0,0}|E_Z}{\hbar(\hbar\omega_0)^2} 
    \cos\phi_{s,s} = \Omega_0 \cos\phi_{s,s}. \label{Omega0}
\end{equation}
Here, $\Omega_0$ is the conventional EDSR Rabi frequency, as observed in material platforms without valleys \cite{Rashba2008}. 
However, Eq.~(\ref{Omega0}) also contains a valley phase factor, which is \emph{not} present in conventional theories.
This is due to the fact that the Dresselhaus SOC in the conduction band of Si 
is an intervalley process \cite{Woods2023,WoodsGFactor}.
The presence of a valley phase in $\Omega_\text{orb}$ has important consequences for EDSR, since the valley phase is spatially randomized in the presence of alloy disorder, as discussed below.

The second term in Eq.~(\ref{Omega}) arises uniquely from the valley physics:
\begin{equation}
    \Omega_v = 
    \Omega_0
    \frac{|\Delta_{p_y,s}|^2}{|\Delta_{s,s}|^2}
    \sin(\phi_{p_y,s} -\phi_{s,s} ) \sin\phi_{p_y,s}. \label{OmegaDipole}
\end{equation}
The processes contributing to $\Omega_v$ are illustrated in Fig.~\ref{Fig1}(e).
Here, in analogy with Fig.~\ref{Fig1}(c), 
the valley splitting, the spin-valley coupling, and the valley dipole take the place of the orbital splitting, the spin-orbit coupling, and the orbital dipole. 
As noted above, ``valley dipole'' refers to the hybridization of valley states, which allows the electron to shift positions and oscillate in response to an ac driving field.
Such valley-orbit coupling can only occur in the presence of alloy disorder; in the present formalism, it is embodied in the coupling term $\Delta_{p_y,s}$ in Eq.~(\ref{Deltann}).
In Eq.~(\ref{OmegaDipole}), it is important to note that $\Omega_v$ is inversely proportional to $|\Delta_{s,s}|^2$, resulting in a large Rabi frequency when the valley splitting is small. 
Like $\Omega_\text{orb}$, $\Omega_v$ also depends on the valley phase $\phi_{s,s}$, as well as on $\phi_{p_y,s}$. 

To characterize the individual contributions of $\Omega_{\T{orb}}$ and $\Omega_v$ to the total Rabi frequency $\Omega$, we now consider the statistical variations of the intervalley matrix elements, caused by disorder.
Since these matrix elements are complex, we may decompose them into their real and imaginary parts: $\Delta_{n_y,n_y^\prime}=R_{n_y,n_y^\prime}+iI_{n_y,n_y^\prime}$.
Within alloy-disorder models, the spatial variations of $R_{n_y,n_y^\prime}$ and $I_{n_y,n_y^\prime}$ are described by independent random variables, each following a normal distribution \cite{Losert2023, WoodsGFactor}.
Consequently, the magnitudes, $|\Delta_{s,s}|$ and $|\Delta_{p_y,s}|$, follow Rayleigh distributions. 
Furthermore, $\Delta_{s,s}$ and $\Delta_{p_y,s}$ are uncorrelated with each other, due to the $s$ and $p_y$ orbitals having opposite parity.
The valley phases, $\phi_{s,s}$ and $\phi_{p_y,s}$, are uniformly distributed in $[-\pi,\pi]$, and are uncorrelated with $|\Delta_{s,s}|$ or $|\Delta_{p_y,s}|$, or with each other. 
As described in Appendix~\ref{sec:delta_covariance}, the covariance matrix elements are given by 
\begin{gather}
    \left< R_{s,s}^2 \right> = \left< I_{s,s}^2 \right> = \sigma_\Delta^2/2, \label{RssSq} \\
    \left< R_{p_y,s}^2 \right> = \left< I_{p_y,s}^2 \right> = \sigma_\Delta^2/4, \label{RpysSq} \\
    \left< R_{s,s} R_{p_y,s} \right> = \left< I_{s,s} I_{p_y,s} \right> = 0, \label{RssRpys}\\
    \left< R_{n_y,n_y^\prime} I_{n_y'',n_y'''} 
    \right> = 0 ,
    \label{eq_var_delta_mn}
\end{gather}
which fully characterize the corresponding valley-coupling distributions.
Here, the averages $\langle\cdot\rangle$ are taken over dot locations or, equivalently, alloy-disorder realizations.

\begin{figure}[t]
\centering
\includegraphics[width=0.49\textwidth]{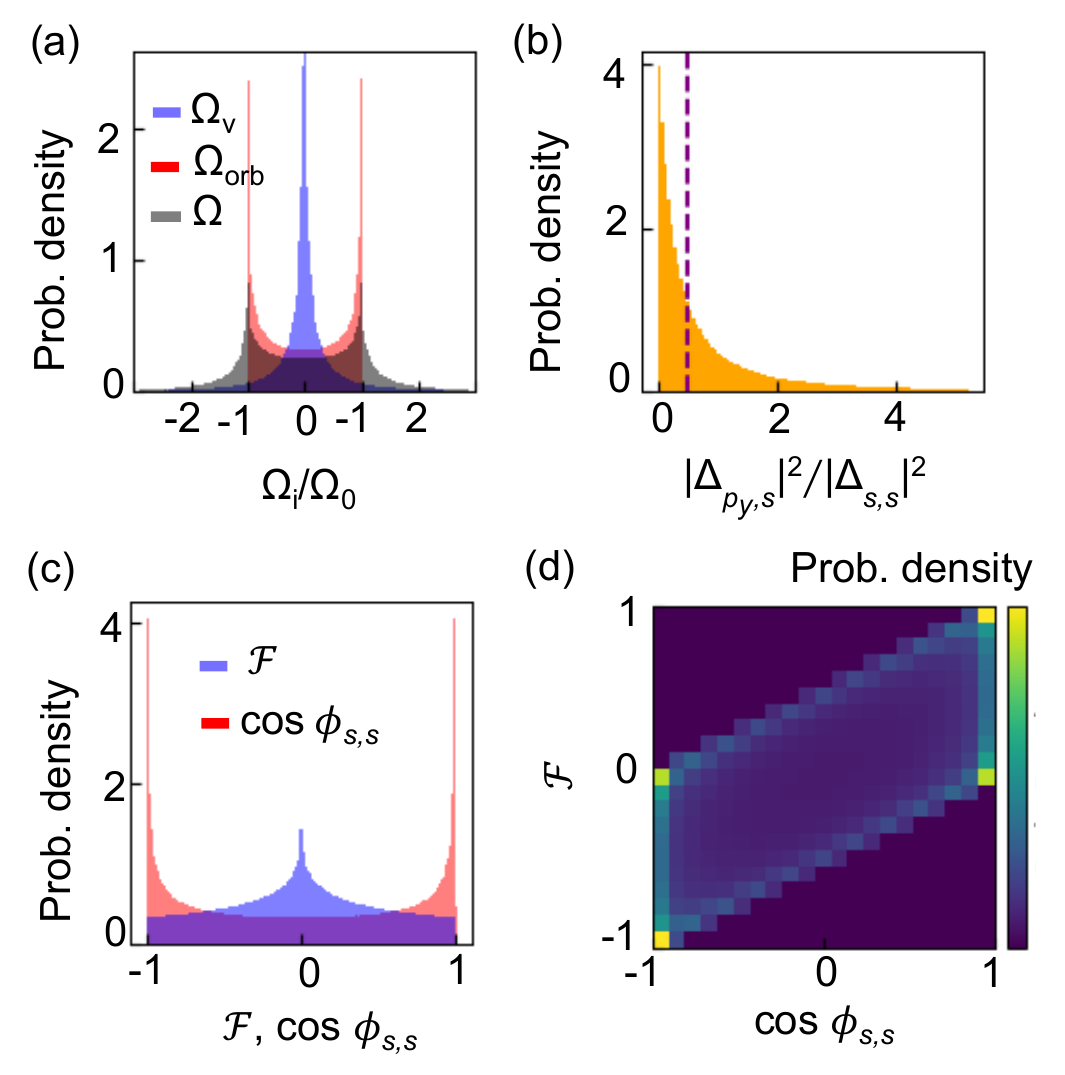}
\caption{
Rabi frequency distributions in the presence of alloy disorder.
(a) Probability distributions of $\Omega_v$, $\Omega_\text{orb}$, and the total Rabi frequency $\Omega=\Omega_v+\Omega_\text{orb}$, all scaled by the conventional EDSR frequency $\Omega_{0}$, defined in Eq.~(\ref{Omega0}). 
(b) Probability distribution of the matrix-element ratio $|\Delta_{p_y, s}|^2/|\Delta_{s, s}|^2$, which contributes to $\Omega_v$ as shown in Eq.~(\ref{OmegaDipole}).
Here, the dashed purple line corresponds to the median value of 0.5. 
(c) Probability distributions of the cosine of the valley phase $\phi_{s,s}$, and valley-phase modulation factor, $\mathcal{F}(\phi_{s,s},\phi_{p_y,s})$, defined in the main text. 
(d) 2D color-scale histogram of $\mathcal{F}$ and $\cos\phi_{s,s}$, showing positive correlations between the two distributions.}
\label{Fig2}
\end{figure}

The distributions of $\Omega_{\T{orb}}$, $\Omega_v$, and $\Omega$ are plotted in Fig.~\ref{Fig2}(a) as red, blue, and gray histograms, respectively. 
Each of these distributions are built from $10^7$ samples of $\Delta_{s,s}$ and $\Delta_{p_y,s}$. 
The distribution of $\Omega_{\T{orb}}/\Omega_0$ is derived from the statistical properties of $\phi_{s,s}$, taking  the form
\begin{equation}
    f_{\Omega_{\T{orb}}/\Omega_0}(x) =
    \begin{cases}
    \frac{1}{\pi \sqrt{1 - x^2}}, & |x| < 1,\\
    0, & |x| \geq 1.
    \end{cases}
\end{equation}
This function is bounded between $\pm 1$ and is concentrated near its extreme values. 
The behavior of $\Omega_v/\Omega_0$ is quite different, with a peak at zero and tails that extend beyond $\pm 1$. 
This behavior primarily arises from the $|\Delta_{p_y,s}|^2/|\Delta_{s,s}|^2$ term in Eq.~(\ref{OmegaDipole}), whose distribution is plotted in Fig.~\ref{Fig2}(b).
We can also compute this distribution analytically, using the Jacobian theorem for transformations with multiple variables \cite{Casella2002}, yielding
\begin{equation}
    f_{|\Delta_{p_y,s}|^2/|\Delta_{s,s}|^2}(x) = \frac{2}{(2x+1)^2} .
\end{equation}
This distribution is peaked at zero and has a median value of $1/2$. 
The valley phase factor appearing in Eq.~(\ref{OmegaDipole}), defined as $\mathcal{F}(\phi_{s,s},\phi_{p_y,s}) = \sin(\phi_{p_y,s} -\phi_{s,s} ) \sin\phi_{p_y,s}$, also contributes to the peaked behavior of $\Omega_v$, as shown in Fig.~\ref{Fig2}(c). 
Although the $\Omega_v$ distribution is concentrated at zero, we see from Fig.~\ref{Fig2}(a) that $\Omega_v$ boosts the average and median values of $|\Omega|$ compared to $|\Omega_\text{orb}|$. 
This is due to positive correlations between the $\cos\phi_{s,s}$ and $\mathcal{F}$ factors in Eqs.~(\ref{Omega0}) and (\ref{OmegaDipole}). 
This correlation is illustrated in Fig.~\ref{Fig2}(d), where we plot a two-dimensional histogram of $\cos\phi_{s,s}$ and $\mathcal{F}$. 
Thus, we conclude that valley-orbit interactions, on average, increase the EDSR Rabi frequency beyond the normal  spin-orbit contributions. 

\section{Rabi frequency spatial fluctuations and dephasing} \label{SpatialFlucsXY}

In this section, we study the statistical variations of the Rabi frequency, caused by alloy disorder, and their effect on qubit dephasing, caused by charge noise.
We show that, although $\Omega$ fluctuates rapidly in space, many ``sweet spots'' arise, where the effects of charge noise are suppressed.

\begin{figure}[t]
\centering
\includegraphics[width=0.49\textwidth]{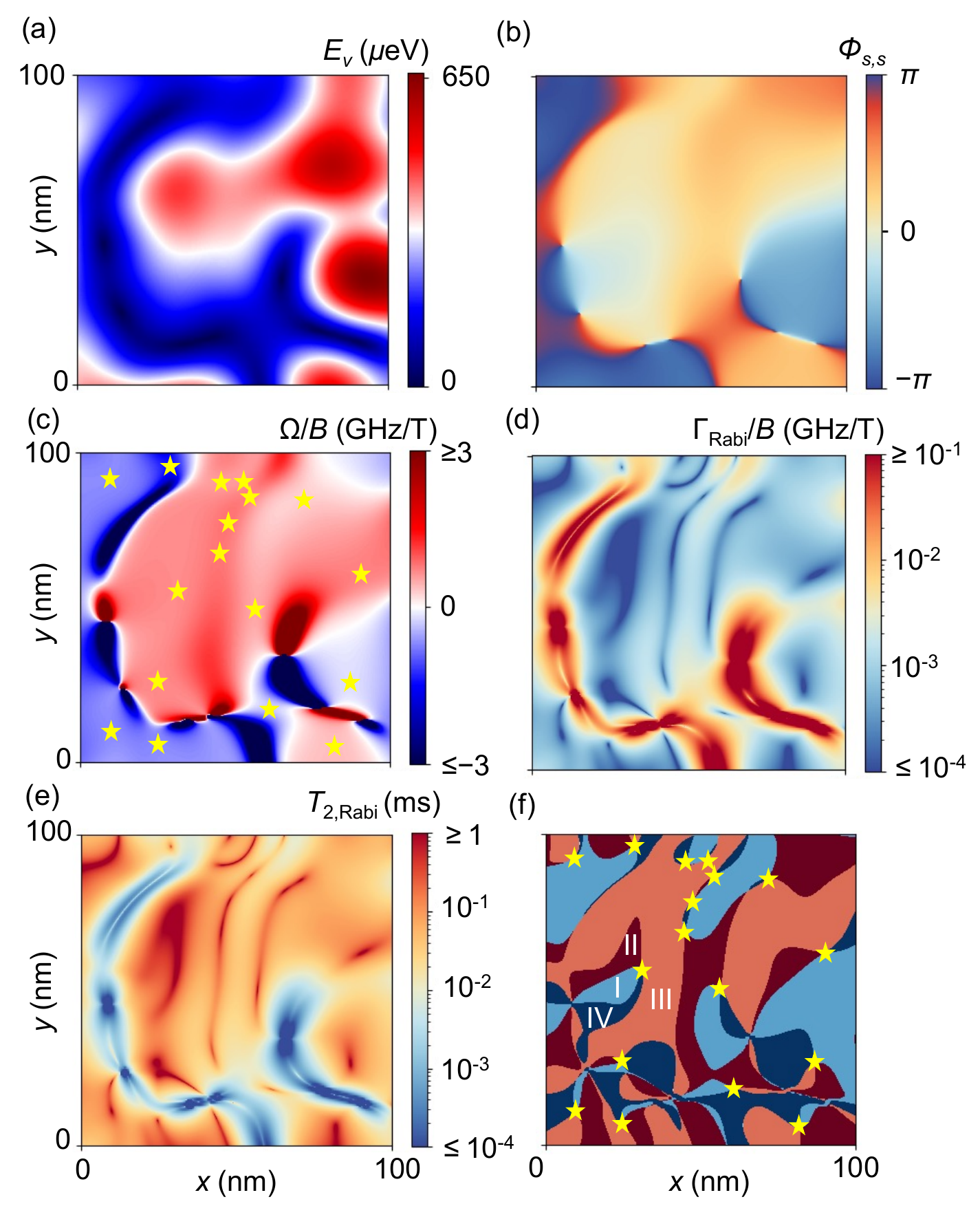}
\caption{
Typical disordered landscapes, for several parameters contributing to EDSR.
(Note that the same random-alloy realization is assumed for all the figures in this work.)
(a) Valley splitting landscape as a function of dot position $(x_d,y_d)$, which we denote as $(x,y)$ here.
(b) The corresponding intervalley phase $\phi_{s,s}$, which exhibits singularities (``valley vortices''), where $E_v=0$, and $\phi_{s,s}$ winds by $\pm 2\pi$. 
(c) The total Rabi frequency $\Omega$, scaled by the magnetic field $B$.
Here, yellow stars indicate sweet spots, defined by $\bm{\nabla}\Omega=0$, where the qubit is relatively insensitive to $\Omega$ fluctuations. 
(d) The scaled Rabi dephasing rate $\Gamma_\text{Rabi}$, arising from $\Omega$ fluctuations. 
Note that the strongest dephasing occurs near valley vortices. 
(e) The Rabi dephasing time $T_{2,\text{Rabi}}=1/\Gamma_\text{Rabi}$, for $B = 20~\text{mT}$.
Note that the best (i.e., longest) dephasing times are observed near sweet spots.
(f) The procedure used to identify sweet spots. 
Here, four regions are identified by colors, according to their signs, with the following labels.
I: $\frac{\partial\Omega}{\partial x}<0,\frac{\partial\Omega}{\partial y}>0$; 
II: $\frac{\partial\Omega}{\partial x}<0,\frac{\partial\Omega}{\partial y}<0$;
III: $\frac{\partial\Omega}{\partial x}>0,\frac{\partial\Omega}{\partial y}<0$;
IV: $\frac{\partial\Omega}{\partial x}>0,\frac{\partial\Omega}{\partial y}>0$. 
The intersection of four colors corresponds to a sweet spot.
}
\label{Fig3}
\end{figure}

Spatial maps of the valley matrix elements $\Delta_{s,s}(x,y)$ and $\Delta_{p_y,s}(x,y)$ are obtained using the covariance function method of Ref.~\cite{WoodsGFactor}, details of which are given in Appendix~\ref{appendix_disorder}. 
This approach also yields maps of the valley phases, $\phi_{s,s}(x,y)$ and $\phi_{p_y,s}(x,y)$.
An example map of the valley splitting $E_v(x,y)=2|\Delta_{s,s}(x,y)|$, obtained in this way, is shown in Fig.~\ref{Fig3}(a). 
Here, we assume $\ell_t = 14~\T{nm}$, corresponding to an orbital energy splitting of $\hbar \omega_t = 1~\T{meV}$.
We also assume a subband energy splitting of $\hbar \omega_l \approx 10~\T{meV}$, and a disorder strength of $\sigma_{\Delta} = 140~\mu\T{eV}$, consistent with a WW with an average Ge concentration of $n_{\T{WW}} = 5\%$ inside the quantum well \cite{WoodsGFactor} (see Appendix~\ref{appendix_disorder}). 
As consistent with previous work \cite{Losert2023,Lima2023b,WoodsGFactor}, the valley splitting is seen to fluctuate dramatically across the device. 
These fluctuations reveal the presence of valley vortices, which are points where $E_v(x,y) = 0$ \cite{Woods2025b}.  
The valley phase $\phi_{s,s}$ also fluctuates randomly, as reported in Fig.~\ref{Fig3}(b).
Here, the valley vortices are identified as the singular points where the phase winds by $\pm 2\pi$.
Using Eqs.~(\ref{Omega0}) and (\ref{OmegaDipole}), we can also obtain a fluctuating map of the total Rabi frequency $\Omega$, as shown in Fig.~\ref{Fig3}(c).
As discussed in Sec.~\ref{sec: EDSR theory}, $\Omega$ is greatly enhanced near a valley vortex, and more generally, along the lines where $E_v(x,y)=2|\Delta_{s,s}(x,y)|\approx 0$, as consistent with Eq.~(\ref{OmegaDipole}).
Indeed, these regions are responsible for the long tails observed in the $\Omega$ and $\Omega_v$ distributions of Fig.~\ref{Fig2}(a). 
On the other hand, $\Omega$ also exhibits strong gradients, especially near the white lines where $\Omega$ changes sign in Fig.~\ref{Fig3}(c).

Unfortunately, strong gradients in $\Omega$ tend to induce qubit dephasing. 
Indeed, charge noise arising from two-level fluctuators \cite{Connors2019} and gate voltage fluctuations \cite{Wang2024} cause random spatial dot displacements, leading  to random shifts of the Rabi frequency. 
We model these fluctuations here by considering randomized in-plane electric fields, $(\delta F_x,\delta F_y)$, for which both components are assumed to be normally distributed, with a mean value of zero, and with the same standard deviation of $\sigma_{F_x} = 0.11~\mu\T{V/nm}$, as consistent with experimental measurements of the detuning noise $\sigma_{\epsilon} = 11~\mu\T{eV}$ \cite{Watson2018,Kranza2020} in double dots separated by $100~\T{nm}$. 
Electric field fluctuations $\delta F_x$ can be converted into position fluctuations by completing the square in the harmonic confinement potential: $(m_t \omega_0^2/2) x^2 + e\delta F_x x = (m_t \omega_0^2/2)(x + e\delta F_x/m_t\omega_0^2)^2 - (e\delta F_x)^2/(2m_t\omega_0^2)$, yielding the relation $\sigma_x=e\sigma_{F_x}/m_t\omega_0^2= 44~\T{pm}$ (and similarly for $\sigma_y$), where we have assumed an orbital splitting of $\hbar \omega_0 = 1~\T{meV}$. 
In turn, position fluctuations are converted to Rabi-frequency fluctuations via 
\begin{equation}
    \sigma_{\Omega} = \sigma_{x}\sqrt{(\partial_x \Omega)^2 + (\partial_y\Omega)^2}, \label{stdRabi}
\end{equation}
which depend, locally, on the Rabi-frequency landscape.
(Note that here and throughout this work, numerical derivatives are performed by evaluating $\Omega$ at points separated by much less than the dot radius $l_t$.)
The average effect of such fluctuations is to suppress the coherence of Rabi oscillations,
with a time envelope given by $f(t) = \exp\left(-\sigma_\Omega^2 t^2/2\right)$ \cite{Ithier2005}, which defines the Rabi dephasing time $T_{2,\T{Rabi}} = \sqrt{2}/\sigma_{\Omega}$ and dephasing rate $\Gamma_{\text{Rabi}} = \sigma_{\Omega}/\sqrt{2}$. 
Spatial maps of these quantities are plotted in Figs.~\ref{Fig3}(d) and \ref{Fig3}(e), where we choose $B=20~\T{mT}$. 
We see that the results vary over many orders of magnitude and exhibit some interesting features.
For example, regions near valley vortices have large $\Gamma_{\T{Rabi}}$, caused by large $\Omega_v$ gradients.
We also observe fine line-like features of low-$E_v$, connecting the vortices.
While $\Omega_v$ exhibits large gradients, with correspondingly large $\Gamma_\text{Rabi}$ values near these lines, the gradients are strongly suppressed at the $E_v$ minima, yielding sudden dips in $\Gamma_\text{Rabi}$ that explain the white lines.
On the other hand, sweet spots also emerge in regions where $\bm{\nabla}\Omega \approx 0$, causing divergences in $T_{2,\T{Rabi}}$, where the effects of charge noise are strongly suppressed. 
(Note that higher-order spatial derivatives and other dephasing mechanisms would prevent any true divergences of $T_{2,\T{Rabi}}$, although we do not include such effects here.)
Figure~\ref{Fig3}(f) shows a method for locating the positions of the sweet spots.
Here, the four colors (and labels I-IV) represent different combinations of the signs of $\partial_x \Omega$ and $\partial_y \Omega$; yellow stars mark the locations of the sweet spots ($\bm{\nabla}\Omega=0$), where four colors meet.
We provide further discussion of sweet spots in Sec.~\ref{LongitudinalNoise}. 

 \begin{figure}[t]
\centering
\includegraphics[width=1\linewidth]{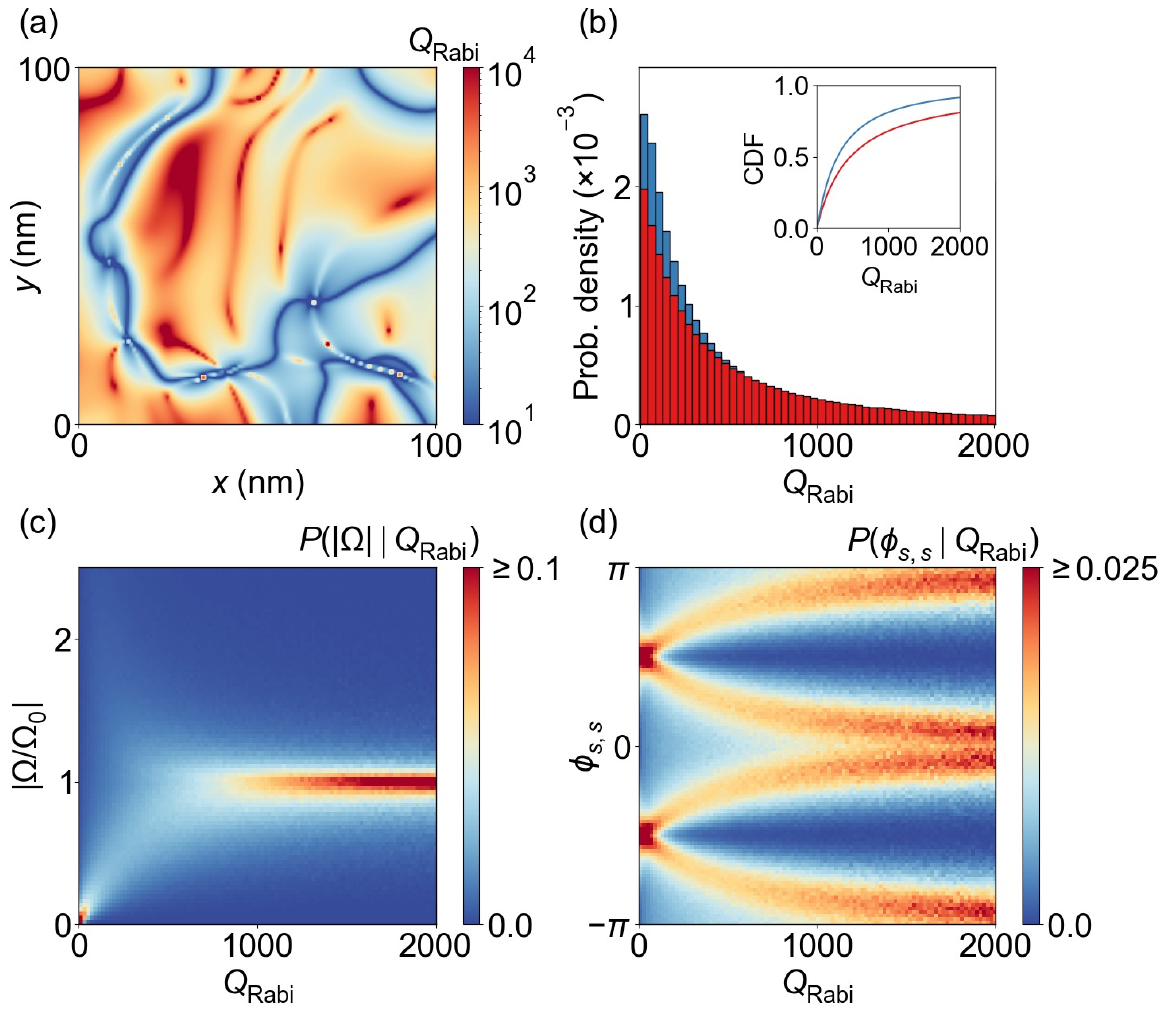}
\caption{Identifying high-fidelity regions of EDSR.
(a) A quality-factor map, $Q_{\text{Rabi}}(x,y)$, based on the same disorder profile as Fig.~\ref{Fig3}, shows a range of behaviors, including high fidelities (dark red, $Q_{\text{Rabi}}\geq10^{4}$), coinciding with sweet spots, and low fidelities (dark blue, $Q_{\text{Rabi}}\leq10$), associated with valley vortices, and regions of high dephasing or low Rabi frequencies. 
(b) Histograms of the quality factors.
Here, red describes contributions from $\Omega_{\text{orb}}$ only, while blue describes the total $Q_{\text{Rabi}}$. 
The inset shows cumulative distribution functions (CDFs) for the same data as the main panel. 
(c) Color-scale plot of the Rabi frequency distribution $|\Omega|$, conditioned on $Q_{\text{Rabi}}$. 
(d) Color-scale plot of the valley-phase distribution $\phi_{s,s}$, conditioned on $Q_{\text{Rabi}}$.} 
\label{Fig4}
\end{figure}

We further characterize the coherence of the EDSR Rabi driving by considering the quality factor $Q_{\T{Rabi}} = \Omega T_{2,\T{Rabi}}$, which quantifies the number of qubit operations that can be performed before decoherence effects dominate. 
Figure~\ref{Fig4}(a) shows a map of $Q_{\T{Rabi}}$ for the same alloy disorder realization used in Fig.~\ref{Fig3}. 
Interestingly, regions of low $Q$ appear to be closely aligned with regions of low $\Omega$ in Fig.~\ref{Fig3}(c), while regions of high $Q$ are aligned with regions of high $T_{2,\text{Rabi}}$ in Fig.~\ref{Fig3}(e).
Generally, the quality factor is highest in regions near the sweet spots, yielding $Q_{\T{Rabi}} > 1000$. 
In contrast, $Q_{\T{Rabi}}$ is suppressed near valley vortices, due to small $T_{2,\T{Rabi}}$ values, despite the fact that $\Omega$ may be high in these regions.

In Fig.~\ref{Fig4}(b), we compare $Q_{\T{Rabi}}$ distributions arising from the total Rabi frequency $\Omega$ (blue) vs the contribution from $\Omega_{\T{orb}}$ only (red), based on $10^{7}$ alloy-disorder realizations. 
Here again we find that, although $\Omega_v$ can diverge near a valley vortex, the $Q_\text{Rabi}$ distribution is dominated by $\Omega_\text{orb}$ contributions.
As we show below, high $Q_\text{Rabi}$ values often occur at locations where $\phi_{s,s}$ takes values of 0 or $\pm\pi$, as suggested by Eq.~(\ref{Omega0}).
(According to the convention we have adopted, this corresponds to the valley phase being aligned or anti-aligned with the spin-orbit phase.)
The cumulative distributions of $Q_{\T{Rabi}}$ are also shown in the inset of Fig.~\ref{Fig4}(b), indicating that very high quality factors, with $Q_{\T{Rabi}} > 1000$, can be expected across 19\% of the spatial map in Fig.~\ref{Fig4}(a).
To explain these high $Q_{\T{Rabi}}$ values, we plot the distributions of $|\Omega|$ and $\phi_{s,s}$, in Figs.~\ref{Fig4}(c) and \ref{Fig4}(d), where both distributions are conditioned on $Q_\text{Rabi}$. 
[Conditional distributions are defined as $P(y|x)=P(x,y)/P(x)$, where $P(x,y)$ is the joint distribution on variables $x$ and $y$, and $P(x)$ is the marginal distribution on $x$.]
In Fig.~\ref{Fig4}(c), for low $Q_{\T{Rabi}}$ values, we observe an interesting (but faint) bimodal distribution concentrated at low Rabi frequencies ($|\Omega| \lesssim  \Omega_0$), and at high frequencies ($|\Omega| \gtrsim \Omega_0$). 
Here, the low Rabi frequencies lead directly to small $Q_{\T{Rabi}}$ values, while the high Rabi frequencies also give low $Q_{\T{Rabi}}$ values, because they originate from valley vortices, near which large frequency gradients enhance the effect of charge noise. 
In contrast, for large $Q_{\T{Rabi}}$, the distributions are concentrated near $|\Omega| = \Omega_{0}$. 
This can be seen from Eq.~(\ref{Omega0}), which gives
\begin{equation}
    \nabla \Omega_{\T{orb}} = -\Omega_0 \sin(\phi_{s,s}) \nabla \phi_{s,s} ,
    \label{NablaOmegaOrb}
\end{equation}
indicating that $\Omega_\text{orb}$ is not strongly affected by charge noise in regions where $\phi_{s,s} \approx 0,\pm\pi$.
Hence, we conclude that high-fidelity EDSR gate operations can be achieved in regions where $\phi_{s,s} \approx 0,\pm \pi$, and away from valley vortices. 
In such regions, $\Omega_v$ contributes insignificantly to $\Omega$, and the remaining contribution, $\Omega_{\T{orb}}$, is largely immune to charge-noise fluctuations. 
These results are further corroborated in Fig.~\ref{Fig4}(d), where we observe the $\phi_{s,s}$ distribution peaked near $0$ and $\pm \pi$, when $Q_{\T{Rabi}}$ is large.  

\section{Impact of out-of-plane electric field noise} \label{LongitudinalNoise}

In Sec.~\ref{SpatialFlucsXY}, we investigated in-plane electric field fluctuations and the corresponding Rabi-frequency fluctuations that lead to dephasing. 
Here, we extend this analysis to include out-of-plane electric field fluctuations. 
We show that the magnitude of the fluctuations can be quite large, and therefore potentially harmful for sweet spots, which were defined above in the absence of vertical fields.
However, we conclude that our solutions for $Q_{\T{Rabi}}$ remain largely robust, and not strongly suppressed, by this noise source. 
We explain this behavior in terms of the spatial structure of $\Omega$.

\begin{figure*}[t]
\centering
\includegraphics[width=0.9\textwidth]{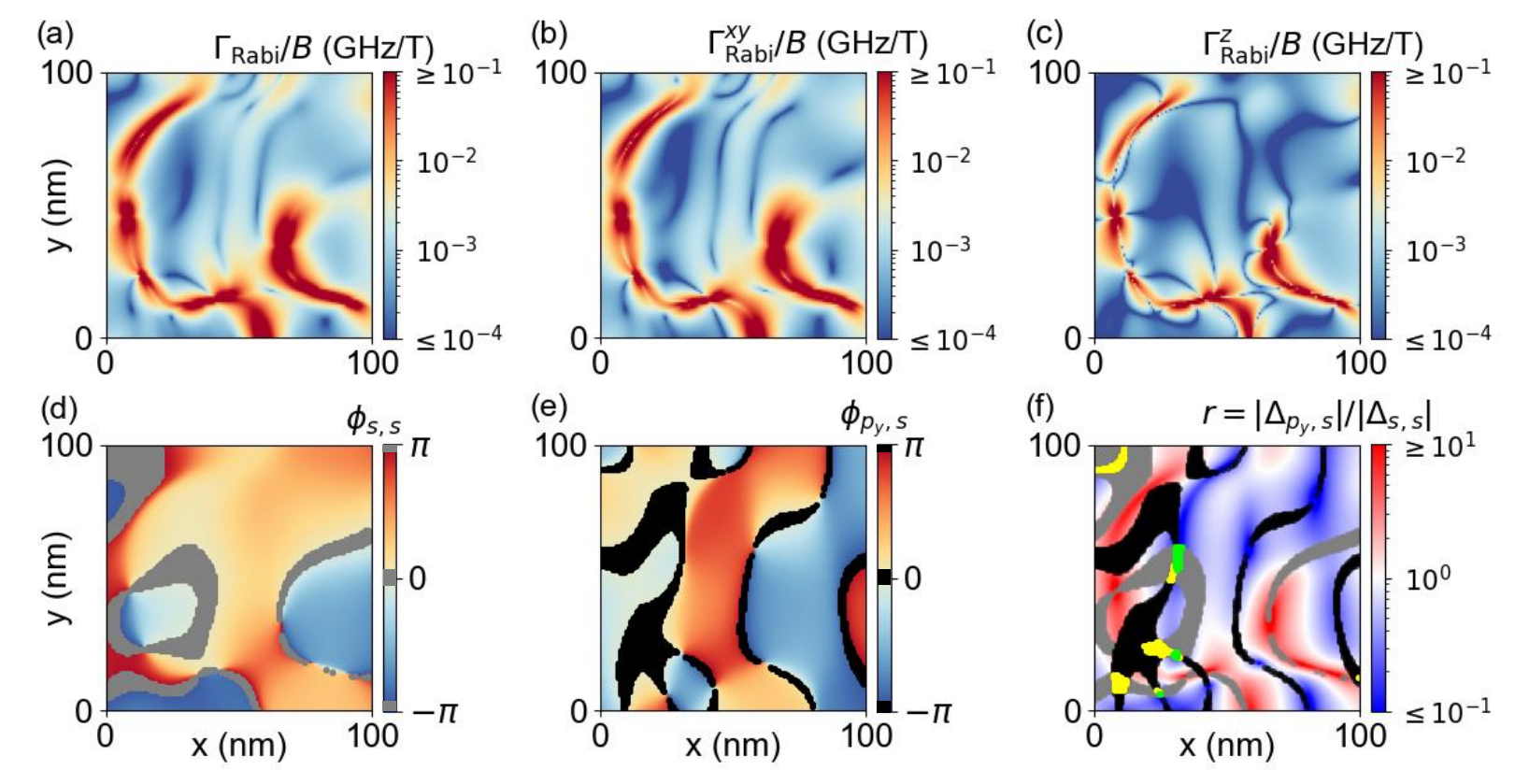}
\caption{Spatial maps of scaled dephasing rates and valley phases ($\phi_{s,s}$ and $\phi_{p_y,s}$), for the same alloy-disorder realization as previous figures. 
(a) The scaled Rabi dephasing rate, including electric field fluctuations in any direction. 
(b), (c) Dephasing-rate maps arising from in-plane and vertical field fluctuations, respectively. Although some low-dephasing regions disappear when both noise components are combined, many remain and are analyzed in panels (d)-(f). 
(d) Spatial distribution of the ground-state valley phase $\phi_{s,s}$.
Gray indicates sweet-spot regions where $|\phi_{s,s}-m\pi|<\epsilon_\text{thresh}$, $m\in\mathbb{Z}$, with $\epsilon_\text{thresh} = 0.045 \pi$. 
(e) Spatial distribution of the higher-orbital valley phase, $\phi_{p_y,s}$.
Black indicates sweet-spot regions where $|\phi_{p_y,s} - n\pi|<\epsilon_\text{thresh}$, $n\in\mathbb{Z}$, with $\epsilon_\text{thresh} = 0.045 \pi$. 
(f) A spatial map of the ratio $r=|\Delta_{p_y,s}|/|\Delta_{s,s}|$, with the gray and black regions of (d) and (e) overlaid. 
As noted in the main text, larger values of $r$ are associated with valley vortices and higher Rabi dephasing rates.
Regions where all the sweet-spot criteria are satisfied, including $r<2$, are indicated with yellow.
Preferred sweet spots with $r<0.1$ are indicated with green.}
\label{Fig5}
\end{figure*}

To explore the effects of vertical field fluctuations, we take a slightly different approach than Sec.~\ref{SpatialFlucsXY}.
There, we assumed that the subband wave functions $\varphi_\nu(z)$ are mainly determined by the confinement potential $V_\text{conf}(z)$ and the vertical electric field $F_z$.
In the current section, we consider a new perturbation (vertical field fluctuations), for which it is not known \emph{a priori} whether the field fluctuations, or the alloy disorder (which also weakly affects the confinement), will have a stronger effect on $\varphi_\nu$.
To answer this question, we include both effects in our calculations, on an equal footing.
We find that this more-rigorous treatment of alloy disorder only weakly affects the overall results, at the 1\% level.
We now describe the full modified calculation.

We extend the previous subband equation~(\ref{SubbandSchro}), as follows:
\begin{equation}
\begin{split}
    \left[-\frac{\hbar^2}{2m_{l}}\frac{d^2}{d z^2} + V_\text{conf}(z) + \mathcal{V}_{s,s}({\bm\rho}_d;z) + eF_z z\right]\varphi_\nu({\bm\rho}_d;z) \\
= \varepsilon_\nu\,\varphi_\nu({\bm\rho}_d;z) .
\end{split}
\label{SubbandSchro2}
\end{equation}
Here, we have explicitly included the disorder potential $\mathcal{V}_{s,s}({\bm\rho}_d;z)\!=\!\mel{ \chi_{s,s}({\bm\rho};{\bm\rho}_d)}{ V_{\text{dis}}({\bm\rho},z)}{ \chi_{s,s}({\bm\rho};{\bm\rho}_d)}$ felt by a dot centered at ${\bm\rho}_d$, as discussed in Appendix~\ref{appendix_disorder}.
In principle, $V_\text{dis}({\bm\rho},z)$ couples the vertical and lateral spatial degrees of freedom; however, the coupling is assumed to be weak, and we incorporate it into Eq.~(\ref{SubbandSchro2}) by projecting it onto the ground-state orbital wave function $\chi_{s,s}({\bm\rho};{\bm\rho}_d)$.
Due to the large subband energy gap, we restrict our attention to the lowest subband, $\nu=0$.

We also include vertical field fluctuations, defined by $F_z=F_{z0}+\delta F_z$, with a typical static field of $F_{z0} = 5~\T{mV/nm}$ that confines the subband wave functions against the top interface of the quantum well.
The random field variable $\delta F_z$ is drawn from a gaussian distribution centered at zero, with a standard deviation of $\sigma_{F_z}$. 
To choose an appropriate value for $\sigma_{F_{z}}$, we consider two noise sources: two-level fluctuators (TLFs) at the semiconductor-oxide interface \cite{Connors2019} and plunger-gate fluctuations \cite{Wang2024}. 
While experiments have not fully characterized these phenomena, evidence suggests that out-of-plane fluctuations may strongly dominate in-plane fluctuations.
To quantify this statement, Ref.~\cite{Wang2024} assumes a noise spectral density of the form $S_{F_z}(f) = A_z^2/f$, where $A_z \approx 3.5~\text{kV}/\text{m}$ is deduced for plunger-gate noise \cite{Xue2022}, including the effects of screening from metal top gates~\cite{Kepa2023}.
The noise variance is then given by
$\sigma_{F_z}^2=\int _{f_\text{min}}^{f_\text{max}} S_{F_z}(f)\,df$, where the integration limits are taken from the experimental bandwidth of Ref.~\cite{Xue2022}, with $f_{\max} = 1/(1~\text{ps})$ representing the time-scale resolution, and $f_{\min} = 1/(8~\text{min})$ representing the measurement duration.
These give $\sigma_{F_z} \approx 20~\mu\text{V}/\text{nm}$, which is nearly 200 times larger than the in-plane fluctuations, $\sigma_{F_x} = 0.11~\mu\T{V}/\text{nm}$, reported above.
Despite this surprising discrepancy, the out-of-plane fluctuations are not expected to overwhelm the in-plane contributions to $\sigma_{\Omega}$ and $Q_{\T{Rabi}}$. 
This is because the total confinement, from the quantum well and the vertical field, strongly suppresses the vertical motion of the dot, giving a standard deviation of $\sigma_{z} = 4.8~\T{pm}$, which is almost an order of magnitude smaller than our previous estimate for the lateral motion: $\sigma_x = \sigma_y = 44~\T{pm}$.
Finally, we consider TLF contributions to the out-of-plane noise.
In Appendix~\ref{appendix_quasi_charge_noise}, we study TLFs of arbitrary orientation, finding only small anisotropies between out-of-plane and in-plane fluctuations.
Combining this with the small size of in-plane fluctuations ($\sigma_{F_x}$), we conclude that out-of-plane fluctuations from TLFs should be overwhelmed by plunger-gate fluctuations.
In the analysis below, we therefore consider only plunger-gate fluctuations of magnitude $A_z = 3.5~\text{kV}/\text{m}$, although it has been noted that this estimate could be on the high side \cite{Kepa2023, Wang2024}.

We now combine these different effects to compute the Rabi frequency. 
The fluctuating field $F_z$ induces time-dependent fluctuations of the vertical wave function $\varphi_0(z)$, which affect the Rabi frequency $\Omega$ in Eqs.~(\ref{Omega})-(\ref{OmegaDipole}) through the coupling parameters $\Delta_{s,s}$, $\Delta_{p_y,s}$, and $\beta_{0,0}$ [Eqs.~(\ref{Deltann}) and (\ref{beta00})].
These quantities all vary spatially, through the disorder potential $V_\text{dis}(\mathbf{r})$.
To analyze the full effect of the fluctuations, we consider one spatial location, and treat $\delta F_z$ as a quasi-static perturbation, sampled from a gaussian distribution; from these chosen parameters, we compute $\Omega$. 
We then repeat this calculation at the same site, for 200 different random $\delta F_z$ values, obtaining an estimate for the standard deviation $\sigma_{\Omega_z}$ at that site.
(The $z$ subscript denotes variations arising from vertical field fluctuations.)
We then repeat this set of calculations as function of position, obtaining $\sigma_{\Omega_z}({\bm\rho}_d)$.
Extending Eq.~(\ref{stdRabi}) to include both vertical and lateral field fluctuations, we finally obtain the full standard deviation of the Rabi frequency due to electric field fluctuations:
\begin{equation}
    \sigma_{\Omega} = \sqrt{\sigma_{x}^2\left[(\partial_x \Omega)^2 + (\partial_y\Omega)^2\right] + \sigma_{\Omega_{z}}^2}. \label{stdRabi2}
\end{equation}
Using this expression, we then compute the spatial variations of $\Gamma_\text{Rabi}$, $T_{2,\text{Rabi}}$, and $Q_\text{Rabi}$, similarly to Sec.~\ref{SpatialFlucsXY}.
The individual contributions from vertical and lateral fluctuations are obtained by setting the appropriate terms in Eq.~(\ref{stdRabi2}) to zero.

We first explore the effects of vertical field fluctuations on sweet spots by computing the total dephasing rate $\Gamma_\text{Rabi}$, including both in-plane and out-of-plane contributions, for the same disorder realization used in Fig.~\ref{Fig3}.
The results are shown in Fig.~\ref{Fig5}(a); for comparison, we also show the contributions from just in-plane fluctuations [Fig.~\ref{Fig5}(b), which is identical to Fig.~\ref{Fig3}(d)], and from just out-of-plane fluctuations [Fig.~\ref{Fig5}(c)].
We note that $\Gamma^{xy}_{\text{Rabi}}$ and $\Gamma^{z}_{\text{Rabi}}$ share the same range here, indicating that both effects contribute significantly and similarly to the dephasing. 
Interestingly, we see that $\Gamma^{xy}_{\text{Rabi}}$ and $\Gamma^{z}_{\text{Rabi}}$ share a similar spatial structure, and that the previously observed sweet spots survive largely intact, in the presence of vertical field fluctuations.
We now explore this important result in further detail.

To understand the common spatial structure of $\Gamma^{xy}_{\text{Rabi}}$ and $\Gamma^{z}_{\text{Rabi}}$, and the robustness of the sweet spots, we consider the first-order variations of $\Omega$, as inferred from Eqs.~(\ref{Omega})-(\ref{OmegaDipole}).
Defining $r=|\Delta_{p_y, s}|/|\Delta_{s,s}|$, we can express these variations as
\begin{align}
        \delta\Omega\propto\, &\sin(\phi_{s,s}) \, \delta \phi_{s,s} \nonumber\\ 
        &+2r \sin(\phi_{p_y, s}-\phi_{s,s})\sin(\phi_{p_y, s}) \, \delta r \nonumber\\ 
&+ r^2  \sin(\phi_{p_y, s}-\phi_{s,s})\cos(\phi_{p_y, s}) \, \delta \phi_{p_y, s}\nonumber\\
&+r^2\cos(\phi_{p_y, s}-\phi_{s,s})\sin(\phi_{p_y, s})[\delta \phi_{p_y, s} - \delta\phi_{s,s} ] .
\label{VarOmega}
\end{align}
Here, we note that $\delta\Omega$ vanishes under the conditions of
$\phi_{s,s} = m\pi$ and $\phi_{p_y,s} = n\pi$, when $m,n \in \mathbb{Z}$.
Each of these conditions describes a curve in the $(x_d,y_d)$ plane, and thus, the sweet spots occur whenever two of these lines cross.
Since the valley phases $\phi_{s,s}$ and $\phi_{p_y,s}$ are smoothly varying across most of the heterostructure, such sweet spots are generally stable and robust.
However, sweet spots also occur near valley vortices, where
the valley phase winds by $\pm 2\pi$, allowing sweet-spot criteria to be satisfied. 
In this case, $r$ is very large, yielding tiny sweet spots that are susceptible to charge-noise perturbations.
To avoid classifying these points as viable sweet spots, we modify our definition of a sweet spot to include the requirement $r < 2$. 
Thus, we identify a set of \emph{practical} sweet spots that satisfy the conditions $r < 2$, $|\phi_{s,s} - m\pi| < \epsilon_{\T{thresh}}$, and $|\phi_{p_y,s} - n\pi| < \epsilon_{\T{thresh}}$, where $\epsilon_{\T{thresh}}$ is a small threshold parameter. 
Fortunately, these criteria are often satisfied in WWs, as demonstrated below, resulting in stable regions of low dephasing, regardless of the direction of the electric field fluctuations. 

In Figs.~\ref{Fig5}(d)-\ref{Fig5}(f), we map out the emergence of sweet spots, based on these criteria.
Figures~\ref{Fig5}(d) and \ref{Fig5}(e) show spatial maps of $\phi_{s,s}$ and $\phi_{p_y,s}$, respectively.
The gray and black regions in these panels denote regions where $|\phi_{s,s} - m\pi| < \epsilon_{\T{thresh}}$ and $|\phi_{p_y,s} - n\pi| < \epsilon_{\T{thresh}}$, respectively, for the threshold value $\epsilon_{\T{thresh}} = 0.045\pi$. 
The same gray and black regions are reproduced in Fig.~\ref{Fig5}(f), atop a spatial map of $r=|\Delta_{p_y, s}|/|\Delta_{s,s}|$. 
Regions where all the sweet spot criteria are satisfied are indicated with yellow. 
As expected, these locations coincide with the regions of small $\Gamma_\text{Rabi}$ in Fig.~\ref{Fig5}(a). 
We also identify in green some especially prominent sweet-spots, corresponding to the stricter condition $r < 0.1$, for which the second, third, and fourth terms in Eq.~(\ref{VarOmega}) are strongly suppressed, due to having small $|r|$.

\begin{figure}
\centering
\includegraphics[width=0.35\textwidth]{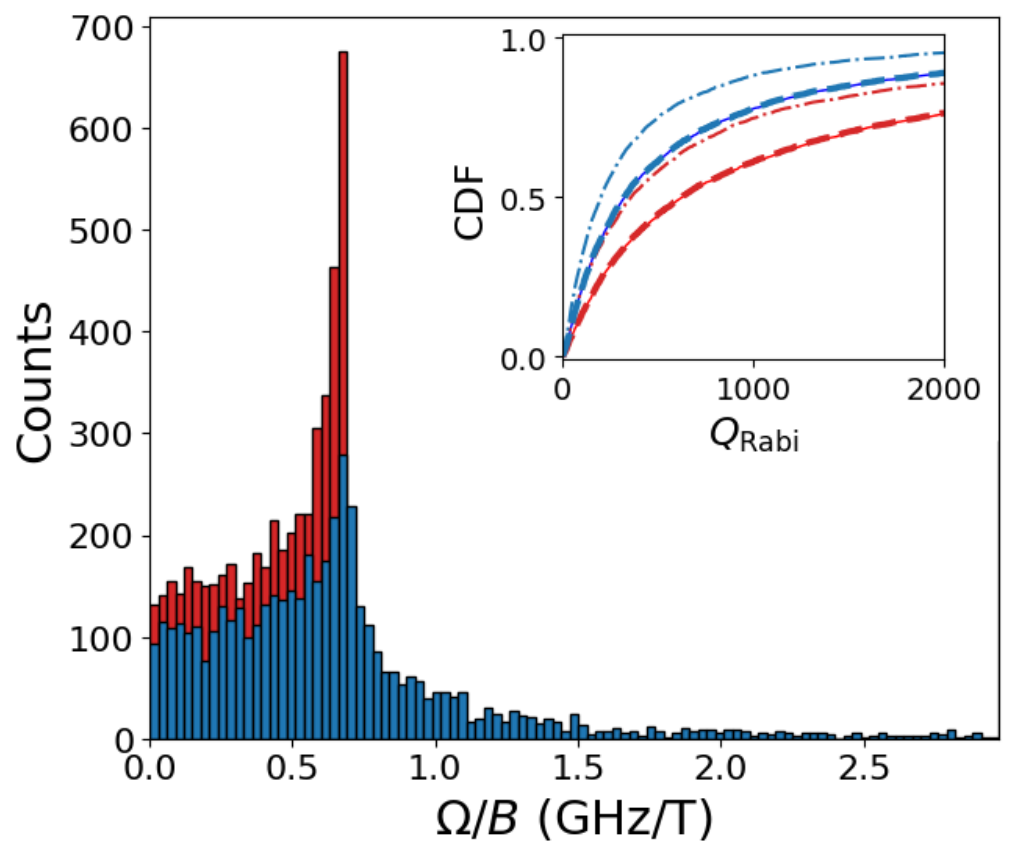}
\caption{Histograms of scaled Rabi frequencies, where $\Omega_\text{orb}$ includes only orbital contributions (red), and $\Omega$ is the total Rabi frequency (blue). 
(For visual clarity, we cut off the plotted range of $\Omega$ values.
The excluded data represent $5.8\%$ of the total samples.)
Note the similarities to Fig.~\ref{Fig2}(a).
The inset shows CDFs for the Rabi quality factor $Q_{\text{Rabi}}$, with colors matching the main panel (red includes only orbital contributions to the Rabi frequency, blue includes the total Rabi frequency).
Solid lines correspond to the case with no vertical field fluctuations ($\delta F_z=0$), closely matching the inset of Fig.~\ref{Fig4}(b).
Dot-dashed lines include the effects of vertical field fluctuations from plunger gates, while dashed lines include fluctuations from TLFs.}
\label{Fig6}
\end{figure}

Finally, we provide a statistical analysis of vertical field fluctuations on $Q_{\T{Rabi}}$. 
For comparison purposes, we sample the $\delta F_z$ fluctuations from two different normal distributions.
The first is defined by $\sigma_{F_{z}} = 20~\mu\T{V/nm}$, corresponding to typical plunger-gate fluctuations, as discussed above. 
The second is defined by $\sigma_{F_{z}} = 0.22~\mu\T{V/nm}$, corresponding to a worst-case scenario for TLF noise.
We estimate this worst-case noise level by scaling the in-plane noise, $\sigma_{F_{x}} = 0.11~\mu\T{V/nm}$, by an upper bound to the fluctuation anisotropy, $\delta F_z/\delta F_{x}\approx 2$, which we determine from the simulations described in Appendix~\ref{appendix_quasi_charge_noise}. 
For both cases, we sample 100 different vertical field values and compute their subband envelopes $\varphi_0(z)$.
We also sample 5,000 random-alloy disorder realizations, while re-computing $\varphi_0$, $\Omega$, and the numerical derivatives of $\Omega$, for each of these 500,000 realizations of random-alloy and vertical-field fluctuations.
The results of these simulations are summarized as histograms in Fig.~\ref{Fig6}. 
The inset also shows the cumulative distribution of $Q_{\T{Rabi}}$ values.
Here, the solid lines are obtained for the case of no vertical electric-field noise, yielding results very similar to Fig.~\ref{Fig4}(d). 
Analogous to Fig.~\ref{Fig4}(d), the red data (both the main panel and the inset) include only orbital contributions to the Rabi frequency ($\Omega_\text{orb}$), while the blue data describe the total Rabi frequency ($\Omega$).
The other curves in the inset include vertical electric-field fluctuations, arising from plunger gates (dot-dashed curves) or TLFs (dashed curves). 
We note that the TLFs have little effect on $Q_{\T{Rabi}}$, compared to the in-plane charge noise, while plunger-gate fluctuations have a stronger effect, reducing $Q_{\T{Rabi}}$ by up to $\sim$50\%. 
Nonetheless, as discussed in Sec.~\ref{SpatialFlucsXY}, we may still achieve large $Q_{\T{Rabi}}$ values by avoiding the regions of low valley splitting.

\section{Conclusion}
\label{conclusion}

In this work, we have assessed the feasibility of using long-period WWs as a platform for implementing high-quality single-qubit gates. 
Due to the presence of Ge throughout the structure, random-alloy disorder becomes a key feature that affects all aspects of EDSR operations.
Here, we have highlighted two main effects.
First, the conventional Rabi driving term ($\Omega_\text{orb}$) is spatially randomized by fluctuations of the valley phase, owing to Dresselhaus SOC being an intervalley process [see Eq.~(\ref{Omega0})].
Second, we have identified an important new Rabi driving mechanism ($\Omega_v$), enabled by the emergence of valley dipoles, which is also randomized by random-alloy disorder [see Eq.~(\ref{OmegaDipole})].
Despite the prominence of disorder in this system, we have demonstrated that fast, high-quality Rabi oscillations can be achieved across most of the heterostructure, without introducing micromagnets.

We have performed simulations of the effects of random-alloy disorder on EDSR using the known statistical properties of valley-coupling fluctuations \cite{Wuetz2021, Losert2023, Woods2023, Lima2023b}, while extending those theories to include valley dipoles.
In this way, we have mapped out the valley splitting and valley phase, and the corresponding Rabi frequency, across typical WW heterostructures.
We have also considered the effects of electric-field fluctuations, both in-plane and out-of-plane, that can reduce EDSR gate fidelities, due to the presence of spatial gradients of the total Rabi frequency, $\Omega=\Omega_\text{orb}+\Omega_v$.
This allows us to derive spatial maps of the coherence time $T_{2,\text{Rabi}}$, which is determined by charge-noise, and the corresponding quality factor $Q_\text{Rabi}=\Omega\, T_{2,\text{Rabi}}$.
These maps allow us to identify and analyze optimal working points or sweet spots, where EDSR gates are relatively insensitive to electric-field fluctuations, in any direction.
Our analytical models show that the sweet spots occur where $\Omega_\text{orb}$ contributions to the Rabi frequency are strong and the valley phase approaches an integer multiple of $n\pi$.
We also identify locations that are unsuitable for EDSR, due to their proximity to valley vortices, where the valley splitting is very small.

Overall, we find that, although valley dipoles suppress the highest quality factors, we can still achieve fast EDSR gate operations and large $Q_{\text{Rabi}}$ factors over most of the spatial landscape.
The first step in implementing high-fidelity gates is then to map out the spatial structure of the intervalley coupling, as demonstrated in recent shuttling experiments \cite{Volmer2024, Volmer2026}.
Our results point to the great potential for using long-period WWs to perform high-quality single-qubit gates, without micromagnets.

\begin{acknowledgments}
This work was supported part by the Army Research Office (Grant No. W911NF-23-1-0110 ). This material is also partially supported by the U.S. Department of Energy, Office of Science, National Quantum Information Science Research Centers as part of the Q-NEXT center. The views and conclusions contained in this document are those of the authors and should not be interpreted as representing the official policies, either expressed or implied, of the Army Research Office (ARO), or the U.S. Government. The U.S. Government is authorized to reproduce and distribute reprints for Government purposes notwithstanding any copyright notation herein.
\end{acknowledgments}

\section*{Data Availability}
The data and simulations that support the findings of this study are available in a Zenodo repository, Ref.~\cite{Zenodo}.

\appendix
\section{Moving-dot transformation}
\label{appendix_a}
One effect of ac driving is to hybridize the ground-state $s$ and $p_y$ orbitals in the laboratory frame. 
This hybridization procedure is simplified in the oscillating frame that follows the bottom of the quantum dot confinement potential, as it is driven by the ac electric field.
In this moving frame, the electron remains in the ground-state $s$ orbital. 
However, matrix elements evaluated in the moving frame then acquire contributions from both $s$ and $p_y$ orbitals in the laboratory frame. In this appendix, we calculate the intervalley matrix elements, $\Delta_{\boldsymbol{n},\boldsymbol{n}^\prime}(t)$, in this moving-dot frame. To perform the calculation, we evaluate Eq.~(\ref{Deltann}) using wave functions in the moving-dot frame. As noted in the main text, we only consider orbital hybridization in the $y$ direction, as consistent with the orientation of ac driving field. 
Hence, in the moving-dot frame where $y\rightarrow y+ y_{0}(t)$, Eq.~(\ref{Deltann}) becomes
\begin{multline}
    \Delta_{n_y,n_y^\prime}(t) = \\
    \int |\varphi_0(z)|^2 
    V_{\T{dis}}(\boldsymbol{r})e^{-i2k_0 z}\chi_{n_y}(y,t)\chi_{n_y'}^*(y,t) dy dz, \label{EqA1}
\end{multline}
where the integration over $x$ makes use of the assumed gaussian wave function.
In Eq.~(\ref{EqA1}), we can write the moving-dot transformation, $\chi_{n_y}(y-y_{0}(t),t)=e^{-i\hat{k}_{y}\cdot y_{0}(t)}\chi_{n_y}(y)$, in terms of the translation operator $\hat{k}_y$, which can be expanded using the ladder operator decomposition, $\hat{k}_{y}=\frac{2}{l_t}(a^{\dagger}_{\hat{y}}-a_{\hat{y}})$, where $l_t = \sqrt{\frac{\hbar}{2m_t\omega_0}}$. Thus, up to leading order in $y_{0}(t)$, we have
\begin{equation}
    \chi_{n_y}(y-y_{0}(t),t)
    \approx\left[1+\frac{l_t}{2}(a^{\dagger}_{\hat{y}}-a_{\hat{y}})y_{0}(t)\right]\chi_{n_y}(y,t) .
    \label{EqA2}
\end{equation}
Inserting this into Eq.~(\ref{EqA1}) and calculating up to first order in $y_{0}(t)$ yields
\begin{eqnarray}
    \Delta_{n_y, n_y'}(t)=&&\Delta_{n_y, n_y'}+\frac{l_t}{2}y_{0}(t)(\sqrt{n_y}\Delta_{n_y-1,n_y'}-\nonumber\\&&\sqrt{n_y+1}\Delta_{n_y+1,n_y'}+\sqrt{n_y'+1}\Delta_{n_y,n_y'+1}\nonumber\\&&-\sqrt{n_y'}\Delta_{n_y,n_y'-1}) .
\label{EqA3}
\end{eqnarray}
Using this result, the intervalley matrix elements of interest for our calculations are given by
\begin{subequations}
    \begin{equation}
        \Delta_{s,s}(t)=\Delta_{s,s}+l_ty_{0}(t)\Delta_{p_{y},s} ,
    \label{EqA4a}
    \end{equation}
    \begin{equation}
        \Delta_{p_{y},s}(t)=\Delta_{p_{y},s}-l_ty_{0}(t)(\Delta_{p_y,p_y}-\Delta_{s,s}).
    \label{EqA4b}
    \end{equation}
\end{subequations}

\section{Schrieffer-Wolff transformations}
\label{appendix_b}
The Schrieffer-Wolff (SW) transformation is a perturbative technique for systematically eliminating couplings to high-energy states through a basis transformation that block diagonalizes the full Hamiltonian. 
This allows one to derive an effective Hamiltonian that can be truncated to include just the low-energy subspace \cite{Winkler2003}. 
As described in the main text, we are interested in the low-energy, 2D spin subspace, corresponding to the ground orbital and valley states. 
Thus, we want to eliminate the couplings to excited orbital and valley states. 
The energy hierarchy for this problem is given by $\hbar\omega_0 \gg E_v \gg E_Z$, which justifies treating couplings to the excited states perturbatively.

We begin by eliminating the orbital excited states, which are well-separated by an orbital energy scale of $\hbar\omega_{0}$ --- the largest energy scale in the problem.
As described in Appendix~\ref{appendix_a}, we consider the moving-dot frame, and recast the Hamiltonian of Eq.~(\ref{HamMatrix}) into the form
\[H=H_{0}+H_{1}+H_{2} .\]
Here, $H_{0}$ is a diagonal matrix containing just the orbital energies, while $H_{1}$ and $H_{2}$ are the perturbations.
We define $H_1$ as the couplings within each orbital subspace ($s$ or $p_y$), including Zeeman energy shifts and intervalley couplings, while $H_2$ includes couplings between the $s$ and $p_y$ orbital subspaces. 
$H_1$ and $H_2$ are therefore given by
\begin{subequations}
    \begin{equation}
        H_{1}=\frac{1}{2}\mu_{B}gB_x\sigma_x\tau_{0}+\Delta_{s,s}e^{i\phi_{s,s}}(t)\tau_{-}+\Delta_{s,s}e^{-i\phi_{s,s}}(t)\tau_{+} 
    \label{EqB1b}
    \end{equation}
    and
    \begin{equation}
        \begin{split}
            H_{2}=&[\Delta_{p_{y},s}(t)+\beta_{0,0}(\hat{k}_{x}\sigma_{x}-\hat{k}_{y}\sigma_{y})]\tau_{-}\\&+[\Delta_{p_{y},s}^{*}(t)+\beta_{0,0}^*(\hat{k}_{x}\sigma_{x}-\hat{k}_{y}\sigma_{y})]\tau_{+}\\&+\frac{(\hbar\omega)eE\sin(\omega t)}{m\omega_{0}^{2}}\hat{k}_{y} ,
        \end{split} 
    \label{EqB1c}
    \end{equation}
\end{subequations}
where $\tau_{0}, \tau_{+},\tau_{-}$ are the identity, creation, and annihilation operators acting on the $\{\ket{+z},\ket{-z}\}$ valley states.

We then perform the Schrieffer-Wolff expansion up to second order and eliminate the high-energy states.
The remaining subspace is 4D, corresponding to the orbital ground state $n_x=n_y=0$, and is spanned by the basis set
$\{\ket{\downarrow,+z},\ket{\downarrow,-z},\ket{\uparrow,+z},\ket{\uparrow,-z}\}$.
The solution is then given by
$\tilde{H}\approx H_{0}+H_{1}+\tilde{H}_{2}$, where $H_0$ and $H_1$ are unchanged, but truncated to the 4D basis set, while
\begin{equation}
    \tilde{H_2}=\begin{pmatrix}
    \zeta(t)&0&i\kappa(t)&ie^{-i\phi_{\beta}}\eta(t)\\
    0& \zeta(t)&ie^{i\phi_{\beta}}\eta(t)&-i\kappa(t)&\\
    -i\kappa(t)&-ie^{-i\phi_{\beta}}\eta(t)&\zeta(t)&0\\
    -ie^{i\phi_{\beta}}\eta(t)& i\kappa(t)&0&\zeta(t)\\
    \end{pmatrix} .
\label{EqB2}
\end{equation}
Here, $\zeta(t)$ is an irrelevant overall energy shift that we henceforth ignore.
We also define $\kappa(t)=-\frac{|\beta_{0,0}||\Delta_{p_{y},s}|}{(2l_t)\hbar\omega_{0}}\sin(\phi_{\beta} - \phi_{p_{y},s})$, and $\eta(t)=-\frac{\hbar\omega eE_\text{ac}|\beta_{0,0}|}{(\hbar\omega_0)^2}\sin(\omega t)$. 
Note that we have made use of the explicit form of the lateral wave functions, which are eigenstates of the parabolic confinement potential describing the quantum dot.

Next, we define a transformation
\[U_{V}=\frac{1}{\sqrt{2}}\begin{pmatrix}
    1&1\\
    -e^{i\phi_{s,s}}&e^{i\phi_{s,s}}
\end{pmatrix}\]
within the valley subspace, which provides a basis transformation from the original basis states $\{\ket{+z},\ket{-z}\}$ to the diagonalized valley states $\{\ket{g},\ket{e}\}$.
However, we recall from Eq.~(\ref{EqA4a}) that $\Delta_{s,s}(t)=\lvert\Delta_{s,s}(t)\rvert e^{i\phi_{s,s}(t)}$ is time-dependent, so that $U_{V}$ is also time-dependent. 
Applying the $U_V$ transformation to $\tilde{H}$ therefore introduces a dynamical correction term of the form $\tilde{\tilde{H}}=U^{\dagger}\tilde{H}U -i\hbar U^{\dagger}\frac{dU}{dt}$, yielding
\begin{widetext}
\begin{equation}
\tilde{\tilde{H}}=\begin{pmatrix}
-\hbar\omega_{Z}-\Delta_{s,s}(t)&-i\varsigma\sin(\omega t)&-\xi(t)&i\kappa(t)+\eta(t)\sin(\phi_{\beta}-\phi_{s,s})\\
i\varsigma\sin(\omega t)&\hbar\omega_{Z}-\Delta_{s,s}(t)&-i\kappa(t)-\eta(t)\sin(\phi_{\beta}-\phi_{s,s})&-\xi(t)&\\
-\xi(t)&i\kappa(t)-\eta(t)\sin(\phi_{\beta}-\phi_{s,s})&-\hbar\omega_{Z}+\Delta_{s,s}(t)&i\varsigma\sin(\omega t)\\
-i\kappa(t)+\eta(t)\sin(\phi_{\beta}-\phi_{s,s})& -\xi(t)&-i\varsigma\sin(\omega t)&\hbar\omega_{Z}+\Delta_{s,s}(t)\\
\end{pmatrix},
\label{EqB3}
\end{equation}
\end{widetext}
where we define $\varsigma=-\frac{eE_\text{ac}\hbar\omega|\beta_{0,0}|}{(\hbar\omega_{0})^2}\cos(\phi_{\beta}-\phi_{s,s})$, $\xi(t)=\frac{\hbar}{2}\frac{d\phi_{s,s}(t)}{dt}$, and $\omega_Z=g\mu_{B}B/2\hbar$. 
Using $\phi_{s,s}(t)= \arctan  \frac{\mathrm{Im}[\Delta_{s,s}(t)]}{\mathrm{Re}[\Delta_{s,s}(t)]}$, where $\Delta_{s,s}(t)$ is defined in Eq.~(\ref{EqA4a}), and expanding up to first order in $|\Delta_{p_y,s}|/|\Delta_{s,s}|$, we obtain $\xi(t) = -\frac{\hbar\omega eE_{ac}}{2m_t\omega_0^2 l_t}\frac{|\Delta{p_y,s}|}{|\Delta{s,s}|}\sin{\omega t} \sin(\phi_{p_y,s} - \phi_{s,s})$.
Here, we note that, in the absence of valley-orbit coupling, we have $\kappa(t)=\xi(t)=0$.
Thus, if we also treat $\eta(t)$ as a higher-order perturbation, Eq.~(\ref{EqB3}) becomes block-diagonal.
In this limit, we recover the following result from Ref.~\cite{Woods2023}: $\mathbf{B}_\text{eff}(t)=\frac{\tilde{\beta}_{0,0}eE_\text{ac}B}{(\hbar\omega_{0})^2}\sin(\omega t)\hat{y}$, where $\tilde{\beta}_{0,0}$ is defined to include the phase modulation of the effective spin-orbit coefficient: $\tilde{\beta}_{0,0}=|\beta_{0,0}|\cos(\phi_{\beta}-\phi_{s,s})$.

We now perform a second SW transformation into the low-energy, 2D subspace, corresponding to the valley ground state $\ket{g}$, making use of the relation $E_v \gg E_Z$.
$\tilde{\tilde{H}}$ from Eq.~(\ref{EqB3}) can now be written as
\[\tilde{\tilde{H}}=\tilde{\tilde{H}}_{0}+\tilde{\tilde{H}}_{1}+\tilde{\tilde{H}}_{2},\]
where $\tilde{\tilde{H}}_{0}$ comprises the diagonal components of Eq.~(\ref{EqB3}), $\tilde{\tilde{H}}_{1}$ describes the intravalley interactions, and $\tilde{\tilde{H}}_{2}$ describes the intervalley interactions. 

Performing the SW expansion up to second order, we obtain $\bar{H}\approx\tilde{\tilde{H}}_{0}+\tilde{\tilde{H}}_{1}+\bar{H_2}$.
Truncating these results to the desired 2D basis $\{\ket{\downarrow},\ket{\uparrow}\}$ gives
\begin{equation}
    \bar{H}_{2}=-\frac{1}{2\Delta_{s,s}(t)}\begin{pmatrix}
    \gamma^2(t)&-2i\xi(t)\kappa(t)\\
    2i\xi(t)\kappa(t)&\gamma^2(t)
    \end{pmatrix},
    \label{EqB4}
\end{equation}
where we define $\gamma^2(t)=\xi^{2}(t)+\kappa^{2}(t)+\eta^{2}(t)$.
Dropping an overall energy shift, the effective dynamics of the spin-$\frac{1}{2}$ electron in the valley ground state is described by
\begin{eqnarray}
    \bar{H}=-\hbar\omega_{Z}\sigma_{x}-\left(\varsigma\sin(\omega t)+\frac{\xi(t)\kappa(t)}{\Delta_{s,s}(t)}\right)\sigma_{y}.
    \label{EqB5}
\end{eqnarray}

To determine the Rabi frequency, we define the interaction picture Hamiltonian, $H_{I}(t)=e^{iH'_{0}(t)}V'(t)e^{-iH'_{0}(t)}$, where $H'_{0}$ is the time-independent part of Eq.~(\ref{EqB5}) and $V'(t)$ contains the time dependence. 
Applying the rotating-wave approximation at the resonance condition $\hbar\omega=g\mu_{B}B$, and expanding $\frac{\xi(t)\kappa(t)}{\Delta_{s,s}(t)}$ up to first order in $y_{0}(t)$, the Rabi frequency is found to be
\begin{multline}
    \Omega=\frac{|\beta_{0,0}|\hbar\omega eE_{ac}}{(\hbar\omega_{0})^2}\cos(\phi_{\beta}-\phi_{s,s})-\frac{|\beta_{0,0}|eE_{ac}\hbar\omega}{(\hbar\omega_{0})^2}\frac{|\Delta_{p_{y},s}|^2}{|\Delta_{s,s}|^2}\\
    \times\sin(\phi_{p_{y},s}-\phi_{s,s})\sin(\phi_{\beta}-\phi_{p_{y},s}).
    \label{EqB6}
\end{multline}
To check the validity of this result, we use Floquet theory to numerically solve Eqs.~(\ref{Eq1b})-(\ref{HV}) of the main text, as described in the following Appendix.

\label{appendix_b}
\section{Alternative calculation of the Rabi frequency using Floquet Theory}
To verify the analytic result of Eq.~(\ref{EqB6}), we perform an independent numerical check of the Rabi frequency using Floquet theory~\cite{Eckardt2015}. 
To implement Floquet theory, we decompose the full Hamiltonian of Eqs.~(\ref{Eq1b})-(\ref{HV}) into its static and dynamic components, expressed as $H(t) = H_0 + H_1 e^{i\omega t} + H_{-1} e^{-i\omega t}$. Here, the dynamic terms represent the alternating electric field, explicitly given by $H_1 = H_{-1}^\dagger = \frac{eE_\text{ac}}{2}\hat{y}$. Floquet theory extends the Hilbert space of the Hamiltonian by introducing ``photon number'' multipliers in units of $\hbar \omega$, corresponding to the driving frequency $\omega$ \cite{Eckardt2015}. Consequently, the extended Floquet Hamiltonian can be represented in block-matrix form as
\begin{equation}
    \label{eq16}
    \begin{pmatrix}
        \ddots & H_{1} & 0 & \\
        H_{-1} & H_{0}+\hbar\omega & H_{1} & 0\\
        & H_{-1} & H_{0} & H_{1} & \\
        & 0 & H_{-1} & H_{0}-\hbar\omega & H_{1}\\
        & & 0 & H_{-1} & \ddots\\        
    \end{pmatrix}.
\end{equation}
Here, we note that each block matrix, $H_{0}$ or $H_{1}$, has dimensions $8\times 8$ (as in the main text), representing the set of low-energy orbital-spin-valley states, ordered by increasing energies in the basis $\ket{n_y, s_i, v_i}$, with $n_x=0$ and $n_y=0,1$.

In the high-frequency limit, where $\hbar\omega \gg \frac{eE_\text{ac}}{2}$, it suffices to consider only one Floquet block in Eq.~(\ref{eq16}), for example, the $16\times 16$ square block including the subblocks labeled $H_0+\hbar\omega$, $H_1$, $H_{-1}$, and $H_0$. 
Following the standard Floquet procedure, we numerically solve for the unitary transformation $U_K$ that diagonalizes $H_0$.
The Floquet Hamiltonian is then given by $H_{F}=U_{K}^{\dagger}H_1U_{K}$. The Rabi frequency is finally obtained by identifying the matrix element in $H_F$ that couples the two lowest-energy Floquet states.

\begin{figure}
\centering
\includegraphics[width=0.35\textwidth]{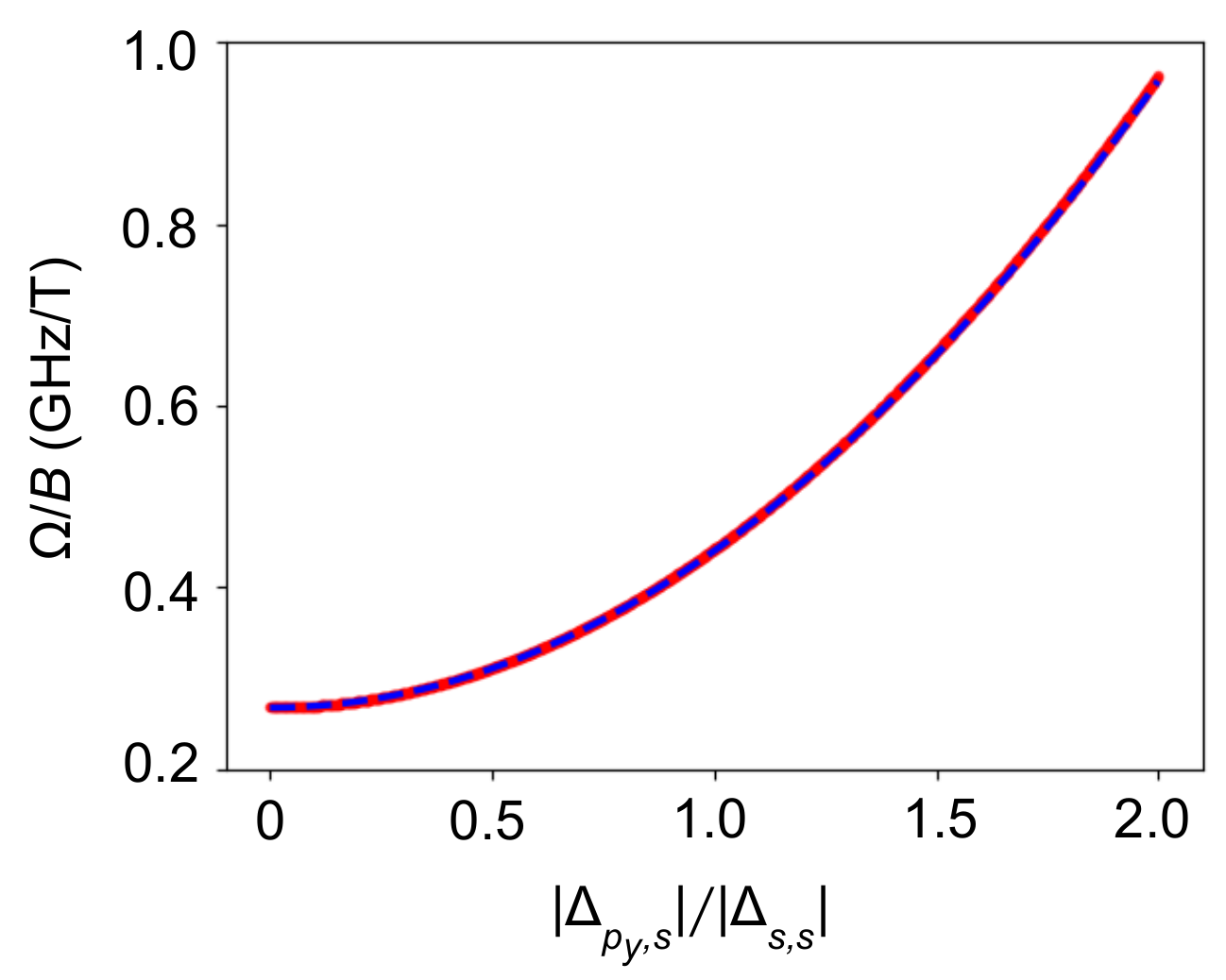}
\caption{Numerical verification of the Rabi frequency formula, Eq.~(\ref{EqB6}). Here, we plot in red the Rabi frequency obtained from Floquet theory, as a function of $\lvert\Delta_{p_y,s}\rvert/\lvert\Delta_{s,s}\rvert$. The analytic relation of Eq.~(\ref{EqB6}) is plotted in dashed blue, showing excellent agreement between the two approaches.}
\label{FigC1}
\end{figure}

To compare our numerical and analytical results, we calculate the Floquet Rabi frequency, as described above, as a function of the ratio $\lvert\Delta_{p_{y},s}\rvert/\lvert\Delta_{s,s}\rvert$, using the parameter values $\phi_{\beta}=\pi/9$, $\phi_{p_y,s}=\pi/5$, $\phi_{s,s}=\phi_{p_y,p_y}=3\pi/4$, $\lvert\Delta_{s,s}\rvert=200\,\mu\text{eV}$, $E_\text{ac}=0.01\,\text{mV}/\text{nm}$, $|\beta_{0,0}|=360\,\mu\text{eV\,nm}$, and $\hbar\omega_0=1\,\text{meV}$. The numerically calculated Rabi frequencies (Floquet method) are plotted in red in Fig.~\ref{FigC1}, while the analytic predictions from Eq.~(\ref{EqB6}) are shown in dashed blue. The excellent agreement confirms the validity of our analytical derivation.

\section{Statistics of Alloy-Disorder Matrix Elements}
\label{appendix_disorder}

In this section, we summarize the statistical properties of the alloy-disorder matrix elements, following the discussion in \cite{WoodsGFactor}. Here, we describe the procedure to obtain the statistics of the alloy-disorder matrix elements, projected onto the lateral basis states, for each $z$ value. We employ a tetragonal lattice with longitudinal and transverse lattice constants defined as 
\begin{equation}
    a_l = \frac{a_0}{4} \quad \text{and} \quad a_t = \frac{a_0}{\sqrt{2}} , \nonumber
\end{equation}
where $a_0$ is the cubic lattice constant of Si. In our model, the Si and Ge sites are treated identically, except that the Ge sites possess an additional on-site energy $E_{\text{Ge}} = 0.6$~eV, which yields the desired band offset between atomic layers with different Ge concentrations. The quantum-well orientation and the Ge concentration oscillations are in the growth direction, $\hat{z}$, so that the local Ge concentration $n_{\text{Ge}}$ is a function of $z$ and is defined as the average over a given atomic layer in the $x$-$y$ plane. 

Assuming that each lattice site is occupied independently, a given site at location $\mathbf{r}=(x,y,z)$ is occupied by Ge with probability $n_{\text{Ge}}(z)$ and by Si with probability $1-n_{\text{Ge}}(z)$. Consequently, the disorder potential at each site is modeled as 
\begin{equation}
    V_{\text{dis}}(\mathbf{r}) =
    \begin{cases}
    -\,E_{\text{Ge}}\,n_{\text{Ge}}(z), & \text{with } \text{prob.}=1 - n_{\text{Ge}}(z), \\[6pt]
    E_{\text{Ge}}\,\bigl[1 - n_{\text{Ge}}(z)\bigr], & \text{with }\text{prob.}= n_{\text{Ge}}(z).
    \end{cases}
\end{equation}
Here, the first and second cases correspond to Si and Ge sites, respectively. This construction leads to 
\begin{align}
    \langle V_{\text{dis}}(\mathbf{r}) \rangle &= 0, \\
    \langle V_{\text{dis}}(\mathbf{r})\,V_{\text{dis}}(\mathbf{r}') \rangle &= \delta_{\mathbf{r},\mathbf{r}'}\,E_{\text{Ge}}^2\,n_{\text{Ge}}(z)\Bigl[1-n_{\text{Ge}}(z)\Bigr].
    \label{Vdisvar}
\end{align}

\subsection{Projecting onto the lateral basis states}
We now consider discretized versions of the quantum dot wave functions, evaluated at the points $({\bm\rho}_i,z_j)$, where ${\bm\rho}_i=(x_i,y_i)$ is a lateral grid point, and $z_j$ is the position of atomic layer $j$.
We define the alloy disorder matrix element at \(z_j\) as
\begin{equation}
    \mathcal{V}_{\bm n,\bm n'}({\bm\rho}_d;z_j) 
    = \sum_{i} V_{\text{dis}}({\bm\rho}_i,z_j)\bar\chi_{\bm n}({\bm\rho}_i;{\bm\rho}_d) \bar\chi_{\bm n'}({\bm\rho}_i;{\bm\rho}_d) ,
\label{eq:Vnn}
\end{equation}
where \(\bar\chi_{\bm n}({\bm\rho}_i)=a_l\,\chi_{\bm n}({\bm\rho}_i)\) is the dimensionless, discretized version of the continuum basis state \(\chi_{\bm n}({\bm\rho})\), and $\bm{n} = (n_x, n_y)$. 
As can be seen from Eq.~(\ref{Deltann}) of the main text, $\mathcal{V}_{\bm n,\bm n'}$ plays an important role in the definition of the valley matrix elements.
Note that we have explicitly included the center location of the dot in Eq.~(\ref{eq:Vnn}), \({\bm\rho}_d = (x_d,y_d)\), since the disorder potential also depends on the dot position.
Making use of this definition, we can compute the two-point correlation function
\begin{align}
    &\langle\mathcal{V}_{\bm n,\bm n'}({\bm\rho}_1;z_i)\,\mathcal{V}_{\bm m,\bm m'}({\bm\rho}_2;z_j)\rangle =\delta_{ij}W_\Delta(z_j)\int d\bm\rho \nonumber\\
    &\qquad\times\chi_{\bm n}(\bm\rho-{\bm\rho}_1)\chi_{\bm n'}(\bm\rho-{\bm\rho}_1)\chi_{\bm m}(\bm\rho-{\bm\rho}_2)\chi_{\bm m'}(\bm\rho-{\bm\rho}_2)
    \label{twopoint},
\end{align}
where $\bm \rho_{1,2}=(x_{1,2},y_{1,2})$ are the lateral positions of quantum dots 1 and 2, and we define the variance at each atomic layer as
\[
W_\Delta(z_j) = \left(E_{\text{Ge}}\, a_t\right)^2\,n_{\text{Ge}}(z_j)\Bigl[1-n_{\text{Ge}}(z_j)\Bigr],
\]
as consistent with Eq.~(\ref{Vdisvar}).

We may compute the correlation functions in Eq.~(\ref{twopoint}) analytically for the orbital basis states $s$ and $p_y$ used in our analysis, since the dot confinement potential corresponds to a simple harmonic oscillator.
This gives
\begin{align}
    &\left<\mathcal{V}_{s,s}({\bm\rho}_1;z_i) \mathcal{V}_{s,s}({\bm\rho}_2;z_j)\right> = \delta_{ij}W_\Delta(z_j) g({\bm\rho}_1,{\bm\rho}_2), \label{Vss}\\
    &\left<\mathcal{V}_{p_y,s}({\bm\rho}_1;z_i) \mathcal{V}_{p_y,s}({\bm\rho}_2;z_j)\right> = \delta_{ij}W_\Delta(z_j)\nonumber \\
    &\hspace{0.9in}  \times g({\bm\rho}_1, {\bm\rho}_2) \left[\frac{1}{2} - \left(\frac{y_1 - y_2}{2\ell_t} \right)^2\right], \label{Vpypy}\\
    &\left<\mathcal{V}_{s,s}({\bm\rho}_1;z_i) \mathcal{V}_{p_y,s}({\bm\rho}_2;z_j)\right> = 
    \delta_{ij}W_\Delta(z_j) \nonumber \\ 
    &\hspace{1.4in}\times g({\bm\rho}_1,{\bm\rho}_2)\left(\frac{y_1 - y_2}{2\ell_t}\right),\label{Vspy}
\end{align}
where we define the correlation function
\begin{equation}
    g({\bm\rho}_1,{\bm\rho}_2) = \frac{1}{4\pi l_t^2}\exp\left(-\frac{|{\bm\rho}_1 - {\bm\rho}_2|^2}{4\ell_t^2} \right).
    \label{eqn:distance, g(r1,r2)}
\end{equation}

\subsection{Generating spatial maps}

To compute spatial maps of the valley coupling parameters, we must first generate spatial maps of $\mathcal{V}_{s,s}({\bm\rho}_i, z_j)$ and $\mathcal{V}_{p_y,s}({\bm\rho}_i, z_j)$.
Thus, our goal is to determine the disorder matrix elements $\mathcal{V}_{s,s}({\bm\rho}_i, z_j)$ and $\mathcal{V}_{p_y,s}({\bm\rho}_i, z_j)$, evaluated at a discrete set of lateral positions $\{{\bm\rho}_1,\ldots,{\bm\rho}_N\}$, for a given atomic layer $z_j$. 
We collect these still-to-be-determined quantities into a single vector with $2N$ components:
\begin{align}
     {\bf y}^{j}
    =
    \Big(
    &\mathcal{V}_{s,s}({\bm\rho}_1,z_j),\ldots,
    \mathcal{V}_{s,s}({\bm\rho}_N,z_j),\nonumber \\
    &\mathcal{V}_{p_y,s}({\bm\rho}_1,z_j),\ldots,
    \mathcal{V}_{p_y,s}({\bm\rho}_N,z_j)
    \Big)^T  \nonumber \\ 
    & = \Big(\ldots,\mathcal{V}_{s,s}({\bm\rho}_i,z_j),\ldots,\mathcal{V}_{p_y,s}({\bm\rho}_i,z_j) ,\ldots\Big)^T .  \nonumber
\end{align}  
The covariance matrix $C^j\in\mathbb{R}^{2N\times 2N}$ describes the statistical behavior of these quantities, in terms of the two-point covariance functions obtained above:
\begin{equation}
    [C^{j}_{(\alpha a),(\beta b)}] = \left\langle \mathcal{V}_{\alpha}({\bm\rho}_a,z_j)\mathcal{V}_{\beta}({\bm\rho}_b,z_j) \right\rangle,
\end{equation}
where $\alpha,\beta\in\{(s,s),(p_y,s)\}$ and $a,b=1,\ldots,N$ label the lateral positions. In block form, we have
\begin{equation}
    C^j
    =
    \begin{pmatrix}
    C^j_{(s,s),(s,s)} & C^j_{(s,s),(p_y,s)} \\
    C^j_{(p_y,s),(s,s)} & C^j_{(p_y,s),(p_y,s)}
    \end{pmatrix},
\end{equation}
where each block is an \(N\times N\) matrix. The diagonal blocks encode the spatial correlations given in Eqs.~\eqref{Vss}--\eqref{Vpypy}, while the off-diagonal blocks encode the cross-covariance given in Eq.~\eqref{Vspy}. 

Next, for each layer $z_j$, we randomly generate an uncorrelated random-disorder vector of length $2N$, given by $\bm x^j \sim \mathcal{N}(\bm 0, I_{2N})$, which denotes a multivariate normal distribution with zero mean and identity covariance. Each of the entries in $\bm x^j$ correspond to an independent, normalized fluctuation of the quantities $\mathcal{V}_{s,s}({\bm\rho}_i,z_j)$ or $\mathcal{V}_{p_y,s}({\bm\rho}_i,z_j)$, at different lateral positions ${\bm\rho}_i$.
We then compute a matrix square root of the covariance matrix, using the Cholesky decomposition: \begin{equation}
    C^j = LL^T,
\end{equation}
where $L$ denotes a lower-triangular matrix. 
Finally, the appropriately correlated alloy-disorder potential vector is given by 
\begin{equation}
    \bm y^j = L \bm x^j.
\end{equation}
For this construction, we note that
\begin{equation}
    \langle \bm{y}^j{\bm{y}^j}^T \rangle = L\langle \bm{x}^j{\bm{x}^j}^T \rangle L^T = LL^T = C^j,
\end{equation}
as desired.
Thus, disorder realizations generated in this way have the desired covariance structure. The first $N$ entries of ${\bm y}^j$ are reshaped into the map of $\mathcal{V}_{s,s}$, while the remaining $N$ entries are reshaped into the map of $\mathcal{V}_{p_y,s}$. Repeating the procedure independently for each atomic layer produces the full disorder landscape as a function of ${\bm\rho}$ and $z$.

\subsection{Calculating valley-coupling parameters}
\label{sec: numerical details}

Making use of these randomly generated (but correlated) disorder parameters, we can now compute different valley-coupling parameters of interest.
In the effective mass approximation, the Hamiltonian along the growth direction \(z\) is given by
\begin{equation}
\tilde H = \frac{\hbar^2}{2m_{l}}\hat{k}_z^2 + V_\text{conf}(z) + \mathcal{V}_{s,s}(z;\bm \rho_d) + eF_z z,
\label{eqn: subband H}    
\end{equation}
where \(m_{l} = 0.91 m_e\) denotes the longitudinal effective mass, and \(F_z\) is the electric field along \(z\). The confining potential with wiggles is represented by \(V_\text{conf}(z)\), while the alloy-disorder potential is characterized by the matrix element $\mathcal{V}_{s,s}(z;\bm \rho_d)$, described above.
In Sec.~\ref{SpatialFlucsXY}, we proceed by setting $\mathcal{V}_{s,s}(z;\bm \rho_d)$ to zero in Eq.~(\ref{eqn: subband H}), while in Sec.~\ref{LongitudinalNoise}, we include these disorder effects in the calculation of $\varphi_0(z)$.

The quantum-well confinement potential \(V_{\text{conf}}(z)\) is defined in terms of the Ge concentration profile,
\begin{equation}
V_\text{conf}(z) = E_{\text{Ge}}\, n_{\text{Ge}}(z),
\label{eq:Vconf}
\end{equation}
which we model as a sigmoid function \cite{Losert2023}:
\begin{multline}
n_{\text{Ge}}(z) = \bar{n}_{\text{Ge}} \left[ 2 - \frac{1}{1+\exp\left(-\frac{z+d/2}{w/4}\right)}  
\right. \\ \left. 
- \frac{1}{1+\exp\left(\frac{z-d/2}{w/4}\right)} \right] 
+ 2 n_{\text{WW}} \sin^2\left(\frac{\pi z}{\lambda_\text{Ge}}\right).
\label{eq:nGe}
\end{multline}
In this expression, we take \(w=1.9\)~nm as the interface width, \(d=10.86\)~nm as the well width, \(n_{\text{WW}} = 0.05\;(5\%)\) as the average Ge concentration in the WW region, \(\bar{n}_{\text{Ge}}=0.3\) (30\%) as average Ge concentration in the barrier region,  \(\lambda_\text{Ge}=1.57\) nm as the wavelength of the Ge oscillations that maximize the spin-orbit coupling \cite{Woods2023}, and \(E_{\text{Ge}} = 0.6\) eV as the band offset. 
To solve Eq.~(\ref{eqn: subband H}) numerically, we discretize the Hamiltonian into a matrix representation for each atomic layer along the $z$ direction. The momentum operator $\hat{k}_z$ is then implemented using the finite-difference method (FDM) as follows:
\begin{align}
    \hat{k}_z^2 = -\frac{\partial^2}{\partial z^2} &=  -\frac{1}{a_{l}^2}\begin{pmatrix}
        -2 & 1 & 0 & 0 & 0 &... \\ 
        1 & -2 & 1 & 0 & 0 &... \\
        0 & 1 & -2 & 1 & 0 &... \\
        0 & 0 & 1 & -2 & 1 &... \\ 
        \vdots & \vdots & \vdots & \vdots & \vdots \\
    \end{pmatrix}. \nonumber 
\end{align}
Finally, we solve for $\bar{\varphi}_0(z_j)$ in this discreted Schr\"odinger equation, which is analogous to Eq.~\eqref{SubbandSchro2}, at every $\bm\rho_d$ value.
Note that $\bar{\varphi}_0(z_j)$ is the discretized, dimensionless version of $\varphi_0(z)$, defined as $\bar \varphi_{0}(z_j) = \sqrt{a_t} \varphi_{0}(z_j)$, and is normalized such that $\sum_j|\bar{\varphi}_0(z_j)|^2=1$.

Based on these results, we may calculate the valley-coupling and SOC operators as 
\begin{align}
    \Delta_{s,s} &= \sum_{j}\, |\bar{\varphi_{0}}(z_j)|^2(V_\text{conf}(z_j) + \mathcal{V}_{s,s}(z_j))e^{-i2 k_0 z_j} , \nonumber\\ 
    \Delta_{p_y,s} &= \sum_{j}\, |\bar{\varphi_{0}}(z_j)|^2(V_\text{conf}(z_j) + \mathcal{V}_{p_y,s}(z_j))e^{-i2 k_0 z_j} , \nonumber\\ 
    \beta_{0,0}  & = \sum_{j}\, |\bar{\varphi_{0}}(z_j)|^2\beta e^{i2k_1 z_j} .
    \label{eqn: exp values of ops}
\end{align}
For example, the valley-splitting, defined as $E_v = 2|\Delta_{s, s}|$, with electric field $F_{z0} = 5~\T{mV/nm}$, is mapped out in Fig.~\ref{Fig3}(a). Similarly, we estimate $\sigma_{\Omega_z}$ in Eq.~\eqref{stdRabi2} by repeatedly solving the Schr\"odinger equation, for many disorder and electric field realizations.
We note that $V_\text{conf}$ contributes only to the deterministic component of $\lvert\Delta_{\BS{n}, \BS{n'}}\rvert$, which is very small in most cases, typically of order $10^{-4}\;\mu\T{eV}$.
In many of the calculations performed in this work, this term is therefore neglected.

\subsection{Same-site covariance functions for the valley-coupling matrix elements}
\label{sec:delta_covariance}

The previous sections outlined the procedure used to determine the spatial maps of the valley parameters. In this Appendix, we derive the same-site variance and covariance of the real and imaginary components of the intervalley matrix element, $\Delta_{\BS{n},\BS{n'}}$, reported in the main text [Eq.~(\ref{eq_var_delta_mn})]. 

We begin by expressing $\Delta_{\BS{n},\BS{n'}}$ in terms of $\mathcal{V}_{\BS{n}, \BS{n'}}(\bm{\rho}_d; z_j)$, the matrix element of the projected alloy disorder potential defined in Eq.~(\ref{eq:Vnn}), yielding
\begin{align}
    &\lvert\Delta_{\BS{n}, \BS{n'}}({\bm\rho}_d)\rvert 
    \exp\big\{i \phi_{n,n'}(\bm{\rho}_d)\big\} 
    =\nonumber \\
    &\qquad  \sum_{j}\, |\bar{\varphi_{0}}(z_j)|^2  \mathcal{V}_{\BS{n}, \BS{n'}}({\bm\rho}_d;z_j) e^{-i 2k_{0} z_j} .
    \label{E6}
\end{align}
Here again, we note that the term $V_\text{conf}$ in Eq.~(\ref{eqn: exp values of ops}) contributes very little to the integral and has been neglected.

Written in terms of its real and imaginary components, $\Delta_{\BS{n}, \BS{n'}}(\bm{\rho}_d)=R_{\BS{n}, \BS{n'}}(\bm{\rho}_d)+i I_{\BS{n}, \BS{n'}}(\bm{\rho}_d)$, the variance and covariance are given by
\begin{align}
\langle R_{\BS{n}, \BS{n'}}^{2}\rangle=\frac{1}{2}\sum_j |\bar{\varphi}_{0}(z_j)|^{4}\langle\mathcal{V}_{\BS{n}, \BS{n'}}(z_j)^2\rangle
\big[1 + \cos(4k_{0}z_j)\big] ,
\label{R_I_correlators_1}
\\
\langle I_{\BS{n}, \BS{n'}}^{2}\rangle= \frac{1}{2}\sum_{j}  |\bar{\varphi}_{0}(z_j)|^{4}
\langle\mathcal{V}_{\BS{n}, \BS{n'}}(z_j)^2\rangle
\big[1 - \cos(4k_{0}z_j)\big] ,
\label{R_I_correlators_2}
\\
\langle R_{\BS{n}, \BS{n'}}
I_{\BS{n}, \BS{n'}}\rangle=\frac{1}{2}\sum_{j} |\bar{\varphi}_{0}(z_j)|^{4}
\langle\mathcal{V}_{\BS{n}, \BS{n'}}(z_j)^2\rangle
\sin(4k_{0}z_j) .
\label{R_I_correlators_3}
\end{align}
In these expressions, we have suppressed the dependence on $\bm{\rho}_d$ for brevity, since the dot location is fixed. 
Next, we note that the sine and cosine terms in Eqs.~(\ref{R_I_correlators_1})-(\ref{R_I_correlators_3}) oscillate rapidly on the scale of the envelope function, such that their averaged contributions may be neglected. Consequently, we find that
\begin{align}
&\langle R_{\BS{n}, \BS{n'}}^{2}\rangle \approx \langle I_{\BS{n}, \BS{n'}}^{2}\rangle  \approx \frac{1}{2}\sum_{j}  |\bar{\varphi}_{0}(z_j)|^{4}\langle\mathcal{V}_{\BS{n}, \BS{n'}}(z_j)^2\rangle,\\
&\langle R_{\BS{n}, \BS{n'}}I_{\BS{n}, \BS{n'}}\rangle\approx 0.
\end{align}
We also find that
\[
\sigma_\Delta^2 = \sum_{j}  |\bar{\varphi}_{0}(z_j)|^{4}\langle\mathcal{V}_{s,s}(z_j)^2\rangle,
\]
as consistent with Refs.~\cite{Losert2023} and \cite{WoodsGFactor}. Using the parameters given in Appendix.~\ref{sec: numerical details}, we estimate that $\sigma_\Delta \approx 160\;\mu\T{eV}$, which is used throughout this work.

\section{Comparing in-plane and out-of-plane electric-field fluctuations from TLFs}
\label{appendix_quasi_charge_noise}

In the main text, we compare the effects of in-plane and out-of-plane electric-field fluctuations, in Secs.~\ref{SpatialFlucsXY} and \ref{LongitudinalNoise}, respectively.
We also consider two sources of noise: plunger-gate fluctuations and two-level fluctuators (TLFs), which are both mentioned in the literature.
Many questions related to charge-noise distributions in the solid state are still under investigation.
In this Appendix, we study one particular question that informs our analyses in Secs.~\ref{SpatialFlucsXY} and \ref{LongitudinalNoise}.
Specifically, we consider the problem of TLFs, which are often thought to reside at the SiGe/oxide or Si/oxide interfaces at the top of the materials stack \cite{Kepa2023, Connors2019, ct_1fnoise_solidstate}.
Given this well-defined location, we may ask the question of whether in-plane or out-of-plane electric-field fluctuations, arising from TLFs, can be expected to dominate, at the location of the quantum-dot electron.

\begin{figure}[t]
    \centering
    %\vspace{-0.8in}
    \includegraphics[width=1\linewidth]{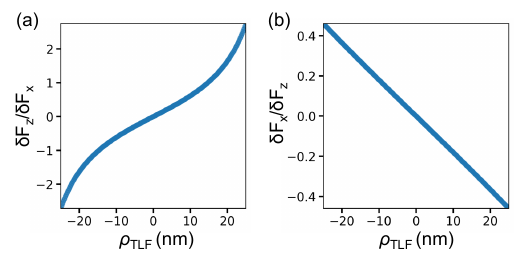}
    \caption{Ratios of in-plane and out-of-plane electric-field fluctuations, caused by a TLF, at the location of a dot electron, as a function of the in-plane vector $\bm{\rho}_\text{TLF}$ that separates the electron and the TLF.
    Results are shown for
    (a) the longitudinal orientation of the TLF (parallel to $\bm{\rho}_\text{TLF}$), and (b) the transverse orientation of the TLF (perpendicular to $\bm{\rho}_\text{TLF}$).}
    \label{fig:tlf}
\end{figure}

We consider a typical experimental materials stack \cite{McJunkin2022, Connors2019}, consisting of (from top to bottom) a metal gate, a 5~nm SiO$_2$ oxide layer, a 60~nm SiGe buffer layer, a 10~nm Si quantum well, and a semi-infinite SiGe virtual substrate.
We adopt values of 3.9, 11.7, and 13.1 for the dielectric constants of SiO$_2$, Si, and the SiGe barrier material, respectively, in units of the vacuum permittivity.
The variations of the dielectric constant inside the WW, due to Ge concentration oscillations, are assumed to be negligible for our purposes.
We model the TLF as a charge trap at the SiGe/SiO$_2$ interface, which fluctuates between two positions, separated by 1~\AA.
We assume the electron in the dot is located 2~nm below the SiGe layer, inside the Si quantum well.
The electron is separated laterally from the fluctuator by an in-plane vector defined as $\bm{\rho}_\text{TLF}$. 
(It is also separated vertically from the TLF by the SiGe buffer layer.)
We consider two different in-plane orientations of the TLF: ``longitudinal,'' meaning parallel to $\bm{\rho}_\text{TLF}$, or ``transverse,'' meaning perpendicular to $\bm{\rho}_\text{TLF}$.

A standard method for solving the electrostatics  problem of charge screening, in the presence of metal and dielectric layers, is the method of images.
For the case of multiple dielectric layers, in principle, an infinite number of image charges is needed.
Here, we adopt an efficient numerical method for solving this problem, as described in Ref.~\cite{ct_comp}: the image charge locations are calculated recursively in each dielectric layer, while satisfying appropriate boundary conditions.
As usual, increasing the number of images provides increasingly accurate answers.
We have compared the converged solution to a simple-dipole solution, where a single image charge of equal magnitude and separation is placed on the opposite side of the screening gate, as commonly assumed in the literature \cite{Wang2024,ct_dp_approx1}.
We find that the full solution, including many images, can differ from the simple-dipole solution by a factor of $\sim$2, although the qualitative behaviors are similar.
Below, we report the quantitatively accurate results, which make use of many image charges.

In Fig.~\ref{fig:tlf}, we plot the ratio of the out-of-plane electric-field fluctuation $\delta F_z$ to the in-plane electric field fluctuation $\delta F_x$, at the location of the electron, as a function of $\rho_{\text{TLF}}$.
The results are shown for (a) the longitudinal orientation of the TLF, and (b) the transverse orientation.
Although we cannot know the exact location of the TLF for a given system, it is clear that both $\delta F_z$ and $\delta F_x$ will decay quickly as a function of $\rho_{\text{TLF}}$.
Therefore, we deduce from Fig.~\ref{fig:tlf} that a reasonable upper bound for the fluctuation anisotropy ratio $\delta F_z/\delta F_x$ should be given by $\sim$2.

\bibliography{apssamp}

\end{document}